\documentclass[a4paper,fleqn,usenatbib]{mnras}
\usepackage[T1]{fontenc}
\usepackage{ae,aecompl}
\usepackage{siunitx}

\usepackage{xcolor}

\usepackage{footnote}

\usepackage{graphicx}	% Including figure files
\usepackage{amsmath}	% Advanced maths commands
\usepackage{amssymb}	% Extra maths symbols
\usepackage{comment}   % for commenting blocks
\usepackage{amsfonts}
\usepackage{booktabs}
\usepackage{siunitx}
\usepackage{hyperref}
\usepackage[normalem]{ulem}
\title[Magnetic field in DR~21]{The magnetic field in the dense photodissociation region of DR~21}

\author[Koley et al.]{Atanu Koley,$^{1,2}$\thanks{E-mail:atanuphysics15@gmail.com}
Nirupam Roy,$^{2}$ %\thanks{Email: oohjh@iisc.ac.in}
Karl M. Menten,$^{3}$
Arshia M. Jacob$^{3}$, 
\newauthor Thushara G. S. Pillai$^{4, 3}$, and Michael R. Rugel$^{3}$
\\
$^{1}$Joint Astronomy Programme, Indian Institute of Science, Bangalore 560012, India\\
$^{2}$Department of Physics, Indian Institute of Science, Bangalore 560012, India\\
$^{3}$Max-Planck-Institut f{\"u}r Radioastronomie, Auf dem H{\"u}gel 69, 53121 Bonn, Germany\\
$^{4}$ Institute for Astrophysical Research, Boston University, Boston, MA 02215, USA
}

\date{Accepted XXX. Received YYY; in original form ZZZ}

\pubyear{2020}

\begin{document}
\label{firstpage}
\pagerange{\pageref{firstpage}--\pageref{lastpage}}
\maketitle

\begin{abstract}

Measuring interstellar magnetic fields is extremely important for understanding their role in different evolutionary stages of interstellar clouds and of star formation. However, detecting the weak field is observationally challenging. We present measurements of the Zeeman effect in the 1665 and 1667~MHz (18~cm) lines of the hydroxyl radical (OH) lines toward the dense photodissociation region (PDR) associated with the compact H{\sc ii} region DR~21~(Main). From the OH 18~cm absorption, observed with the Karl G. Jansky Very Large Array, we find that the line of sight magnetic field in this region is $\sim 0.13$~mG. The same transitions in maser emission toward the neighbouring DR~21(OH) and W~75S-FR1 regions also exhibit the Zeeman splitting. Along with the OH data, we use [C{\sc ii}] 158 $\mu$m line and hydrogen radio recombination line data to constrain the physical conditions and the kinematics of the region. We find the OH column density to be $\sim 3.6\times10^{16}(T_{\rm ex}/25~{\rm K})~{\rm cm}^{-2}$, and that the 1665 and 1667 MHz absorption lines are originating from the gas where OH and C$^+$ are co-existing in the PDR. Under reasonable assumptions, we find the measured magnetic field strength for the PDR to be lower than the value expected from the commonly discussed density--magnetic field relation while the field strength values estimated from the maser emission are roughly consistent with the same. Finally, we compare the magnetic field energy density with the overall energetics of DR~21's PDR and find that, in its current evolutionary stage, the magnetic field is not dynamically important.

\end{abstract}


%%%%%%%%%%%%%%%%%%%%%%%%%%%%%%%%%%%%%%%%%%%%%%%%%%

%%%%%%%%%%%%%%%%% BODY OF PAPER %%%%%%%%%%%%%%%%%%

\section{Introduction}\label{sec:intro}

The  magnetic field, $B$, is expected to play an important role in the dynamics of interstellar clouds in various stages of their evolution and in the formation of stars \citep[for an overview, see ][]{crutcher12}.
Currently, even 
while this is widely accepted as 
plausible,
how and to what extent the magnetic field influences processes acting in the interstellar medium (ISM), is not entirely understood.
Two strikingly different pictures 
have emerged, namely that of magnetically dominated star formation \citep[e.g.][]{mouschovias99, mouschovias91}, 
and a turbulence dominated star formation theory \citep[e.g.][]{maclow04, padoan99}. While the debate is far from being settled, the tremendous progress in magnetohydrodynamic simulations in the last decade start to show that answer lies somewhere in between \citep[for an overview, see ][]{hennebelle2019}. Therefore, observational studies of the magnetic field  strength in molecular clouds in different evolutionary stages are necessary. However, measuring the weak ($\lesssim$ mG) interstellar magnetic field is challenging. Apart from the general issue that $B$-fields in the various components of the ISM are weak, different direct and indirect methods of estimating the magnetic field have their own limitations. For example, the rotation measure, resulting from the Faraday effect, only provides a line of sight density weighted integrated quantity and can only be measured for the ionized ISM. Imaging the linear polarisation of the far infrared or (sub)millimeter radiation of interstellar dust provides the direction of the $B$-field, but not its strength \citep{Hildebrand1988}. \citet{heiles12} give a summary of the various methods used for the measurement of magnetic fields in the different phases of the ISM.

Much of the molecular as well as the atomic neutral ISM of the Milky Way consists of photodissociation regions (PDRs) -- regions in which both energetics and astro-chemistry are regulated by far-ultraviolet (FUV; $\sim 6 - 13.6$ eV) photons. Apart from the rich ISM physics and chemistry involved, on a global scale, PDRs are directly related to the star-gas-star cycle via the formation and destruction of star-forming molecular clouds, and may play a role in regulating the star formation rate via FUV-induced feedback. Now, the two above-mentioned scenarios of star formation make distinct predictions for the spatial gradient of the mass-to-flux ratio ($\frac{M}{\Phi}$) between the core of a molecular cloud and its envelope. In ambipolar diffusion driven fragmentation of magnetically dominated clouds, the quantity $R$ ($\equiv\frac{(M/\Phi)_{\rm core}}{(M/\Phi)_{\rm envelope}}$, i.e. ratio of the mass-to-flux ratio of the core to that of the envelope) is always greater than $1$ \citep[e.g.,][]{tassis07}. In turbulence driven fragmentation, $R$ can have a wide range of values not necessarily larger than $1$ \citep[e.g.,][]{lunttila09}. For this reason, magnetic field measurements to constrain the mass-to-magnetic flux in the core and the PDR envelope of star forming regions, are very promising in distinguishing between the theories for cloud fragmentation.

Zeeman splitting observations of spectral lines are an important tool to directly estimate (often only the line-of-sight component of) the magnetic field in the region where the spectral line is originating from in absorption or emission. There are various spectral lines, with reasonable Zeeman splitting factor, originating from different atomic and molecular species \citep[for a compilation, see Table~1 of][]{Heiles1993}. The target transition can be chosen depending on the nature of the source, these different species and $their$ transitions trace a variety of physical conditions of the ISM. The goal is to relate the results of the magnetic field determinations to the density and temperature, as well as the morphology and kinematics of the sources in different evolutionary stages of interstellar clouds and star formation.

\subsection{The OH radical and its Zeeman splitting}
The hydroxyl radical (OH)  is a tracer of different environments of the molecular ISM and well-suited for magnetic field measurements \citep{crutcher83}. It is widespread in diffuse and translucent as well as dense, dusty interstellar clouds.  
OH is also found in the dense molecular envelopes of (ultra)compact H{\sc ii} regions \citep{Guilloteau1985}, where it forms via the photodissociation of H$_2$O by UV radiation from the newly formed early type stars \citep{elitzur1978, hartquist1991}. In many of the younger \textit{ultra}compact H{\sc ii} regions, OH lines from radio wavelength hyperfine structure (hfs) transitions from a variety of rotational levels show intense maser emission (strongest in the 1665~MHz and 1667~MHz ground-state lines), with the archetypal W3(OH) being the prime example. Some of the lines arising from energy levels high above the ground-state show (enhanced) absorption \citep{Guilloteau1982}. In contrast, these lines exclusively show absorption toward more developed compact H{\sc ii} regions such as DR~21 \citep[e.g.][]{Matthews1986}.

Due to $\Lambda$-doubling and hfs splitting, the ground-state rotational level of the OH ${^2}\Pi_{3/2}$ spin variant splits into a quartet of lines with rest frequencies near $\sim$ 1612, 1665, 1667, 1720~MHz \citep[]{draine11}. These lines are further subject to Zeeman splitting, and  may be used to measure the  magnetic field in molecular clouds where these lines are observed in emission and/or absorption. Please note, if the Zeeman splitting is smaller than the Doppler line width, then the spectral lines are highly blended. Unless the magnetic field is strong enough, such that the splitting is greater than the Doppler width, it is extremely difficult to directly detect the line splitting \citep[]{crutcher93, sault90}. However, as the components of the Zeeman splitting lines are circularly polarised in opposite senses (right and left hand), it is possible to detect the small difference by subtracting the two circular polarizations and looking for the signal in Stokes V. For a magnetic field with a component along the line of sight, $B_{\text{LOS}}$, the relative frequency shift between the right and the left circular polarization is $\delta \nu = \pm z|B_{\text{LOS}}|$, where the Zeeman splitting factors, $z$, are $3.27$ and $1.96$~Hz~$\mu$G$^{-1}$ for the two main lines at 1665~MHz, and 1667~MHz, respectively \citep[]{kazes86,crutcher77}. Hence, the Zeeman splitting measurements of these OH 18~cm lines allow for the  estimation of the magnetic field, even in the relatively low density regime \citep{crutcher93}.On the other hand, OH maser emission arises in compact, high density regions ($n \sim 10^{7}~{\rm cm}^{-3}$), under special circumstances, and the maser lines can be utilised to estimate the field strength in such high density regions. The density of the environment probed in this study, the PDR of DR~21 (Main), lies, at $\sim 10^{6}~{\rm cm}^{-3}$, between the aforementioned values.
 
In this paper, we present observations of the DR~21 molecular cloud complex with the Karl G. Jansky Very Large Array (VLA).  We study the Zeeman splitting of the OH 1665~MHz, and 1667~MHz lines seen in absorption toward the photodissociation regions associated with the H{\sc ii} region DR~21~Main and estimate the magnetic field strength. We also briefly report on maser emission in these lines toward DR~21~(OH). In \S\ref{sec:DR21}, we present a brief overview of the relevant information about the DR~21 cloud. We present the details of the observations and data reduction in \S\ref{sec:observations}. In \S\ref{sec:results}, we describe the method of estimating the magnetic field strength from the Zeeman splitting measurements of the 1665~MHz and 1667~MHz transitions of OH and present our measurements. The results are presented and discussed in \S\ref{sec:discussion} that, to give context to our Zeeman effect determination, in \S\ref{subsec:DR21_PDR} and \ref{subsec:ohabundance}, summarises the physical condition in the  DR~21 (Main) PDR and its OH content. The OH absorption lines are used to estimate the OH column density (\S\ref{subsec:ohcoldens}), to understand the OH kinematics (\S\ref{subsec:ohkinematics}), and, in \S\ref{subsec:energetic} we compute the mass-to-magnetic flux ratio, and then compare  magnetic, thermal and hydrodynamical energy densities to understand the importance of magnetic field for the DR~21~(M) region.
Finally, our main conclusions are summarised in \S\ref{sec:conclusions}.

\section{DR~21}\label{sec:DR21}

\begin{figure}
\includegraphics[width=0.5\textwidth]{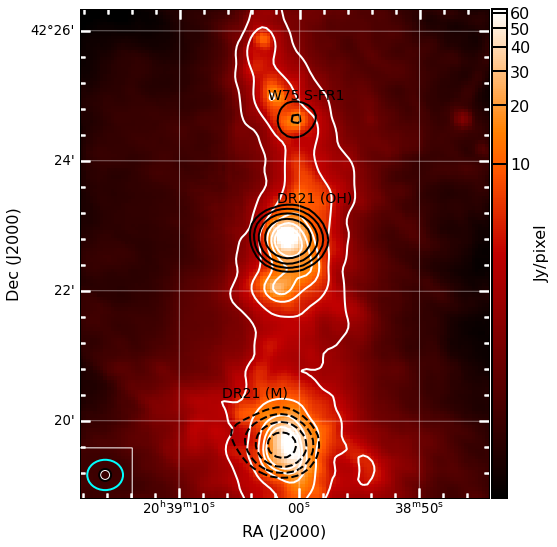}
 \caption{Overview of the DR21 environment showing Herschel 70\,$\mu$m image in color scale \citep{hennemann12}, 1.2\,mm dust continuum emission from MAMBO \citep{Motte2007} in white contours and moment zero map of the OH 1665 MHz transition in black contours from the VLA observation (this work). The 1.2\,mm contours start at 250, 500 mJy in steps of 500 mJy/beam. In the bottom left corner, the Herschel ($9\arcsec.2$), the MAMBO ($11\arcsec$) and the VLA ($33\arcsec\times28\arcsec$) beams are shown in white, black and cyan, respectively.}
 \label{fig:figo}
\end{figure}

The DR~21 molecular cloud is a part of the Cygnus-X massive radio emission/molecular cloud complex, which hosts a large number of young massive stars ranging, in decreasing age, from several OB associations, over H{\sc ii} regions 
to even younger, deeply embedded objects \citep{Reipurth2008}. 
The DR~21 molecular cloud consists of a dense filament, that extends from the core associated with the compact H{\sc ii} region DR~21 (M) near its southern end to the $\sim 3^{\prime}$ (0.4 pc) more northern, younger more deeply embedded  DR~21~(OH) (also known as W~75S)~
\citep{Motte2007, hennemann12}. An overlay of the Herschel 70\,$\mu$m image, 1.2\,mm dust continuum emission from MAMBO \citep{hennemann12,Motte2007} and the OH 1665 MHz transition (VLA, this work) of this region, marking the location of DR~21 (M), DR~21 (OH) and W~75S, is shown in Fig.~\ref{fig:figo}. DR~21~(OH) only shows quite weak radio emission \citep{argon00}, but marks its activity with OH and also with class II CH$_3$OH maser emission \citep{Raimond1969, Menten1991}.
A few other cores in the filament also host maser sources \citep{Motte2007}.
This cloud has been extensively studied at different wavelengths using various molecular tracers like CO, CN, HCN, SiO, CH$_3$OH, HCO$^{+}$, OH, H$_{2}$O, H$_{2}$, H recombination line etc., to name a few \citep[e.g.][]{dobashi19, momjian17, white10, crutcheretal99, garden92, garden90, roelfsema89, wendker84, genzel77, norris82, dickel78}. The intricate filamentary structures of this region, have been mapped in emission from dust \citep[]{hennemann12}, as well as using molecular lines \citep{Schneider2010}. % throughout this complex. 
By astrometric trigonometric parallax measurements the 6.7~GHz CH$_3$OH maser sources associated with DR~21, \citet{rygl12} have established its distance to be $1.50$~kpc (to within 5\% accuracy). Extended mapping in lines of the $^{12}$CO and $^{13}$CO  molecules by \citet{dickel78} and \citet{dobashi19} reveals two major kinematically distinct molecular cloud components, namely the main filamentary DR~21 cloud at $v_{\rm LSR}{\sim -3}$ km~s$^{-1}$ with a mass of ${\sim 31000~}$M$_\odot$, and a ${\sim 9}$ km~s$^{-1}$ component (with a mass of ${\sim 3400~}$M$_\odot$) that represents material at the local standard of rest (LSR) velocity of the extended W~75 
region\footnote{The +9 km~s$^{-1}$ cloud  is sometimes termed the ``W~75 N cloud'', because the faint, compact H {\sc ii} region W~75 N lies within it \citep{dickel78}}. In addition, some intermediate velocity gas is also detected. \citet{dickel78} suggest that the two clouds are colliding, possibly triggering (some of) the observed star formation. This notion is supported by \citet{dobashi19}.

As to magnetic field measurements for sources within the DR~21 filament, previously, based on Zeeman splitting observations of the CN 3~mm lines toward two compact cores in the DR~21~(OH) region (1 and 2), \citet{crutcheretal99} have reported the line of sight component of the magnetic field $B_{\text{LOS}}$ to be $-0.36 \pm 0.10$~mG and $-0.71 \pm 0.12$~mG, respectively. On larger scales, an hourglass-like configuration of the magnetic field in the cloud core around the DR~21~Main (M) H{\sc ii} regions has been proposed based on three dimensional modeling of the observed 350~$\mu$m dust emission by \citet{kirby09}.

In this study, we probe, with OH absorption lines, the magnetic field and kinematics in the photodissociation region (PDR) that forms the interface between the compact H{\sc ii} region DR~21 M and its cooler molecular envelope and compare our findings with earlier results obtained from different tracers.

\begin{table}
	\centering
\caption{Summary of observational parameters.}
	\label{tab:table1}
	\begin{tabular}{lc} 
\hline
VLA proposal ID & 14A-031 \\
Date of Observations & 2014 July 7--August 19 \\
Configuration & D  \\
Observing Band & L (1-2~GHz) \\	
R.A. of field center (J2000) & $20{^{\rm h}} 39{^{\rm m}} 01\rlap{.}^{{\rm s}}0$ \\
Dec. of field center (J2000) & $+42{^\circ} 19^{\prime} 43^{\prime\prime}$ \\
Calibrators & 3C~ 286, 3C~138, \\
~ & J2052$+$3635\\
Continuum:& \\
Observing Bandwidth & $8 \times128$~MHz \\
Number of channels & $8 \times 64$\\
Spectral line: & \\
Rest Frequencies & 1665.40 \& 1667.36~MHz \\
Observing Bandwidth & $2\times 1$~MHz \\
Number of channels & $2 \times 1024$ \\
Channel spacing & 0.175 (km~s$^{-1}$) \\
Primary beam (HPBW) & ${\sim 30^{\prime}}$ \\
Synthesised beam FWHM & ${\sim 33^{\prime\prime}\times 28^{\prime\prime}}$\\
On-source time & ${\sim 7}$~hr~$50$~min \\
RMS noise (per channel)  &${\sim 10}$~mJy~beam$^{-1}$ \\
\hline
	\end{tabular}
\end{table}

\section{Observation and Data analysis}\label{sec:observations}
In \S\ref{subsec:OH18cm} we detail the observational aspects of the OH 18~cm data presented in this work, observed using the VLA. In addition we compare our 18~cm OH spectra with spectra of the $158~\mu$m transition of ionized carbon and  H 72$\alpha$ radio recombination line. Acquisition of the auxiliary data is briefly discussed in \S\ref{subsec:CII} and \S\ref{subsec:HRRL}.
\subsection{Radio wavelength 18~cm data}\label{subsec:OH18cm}
\begin{figure*}
  \includegraphics[width=3.6in,height=3.45in,angle=0]{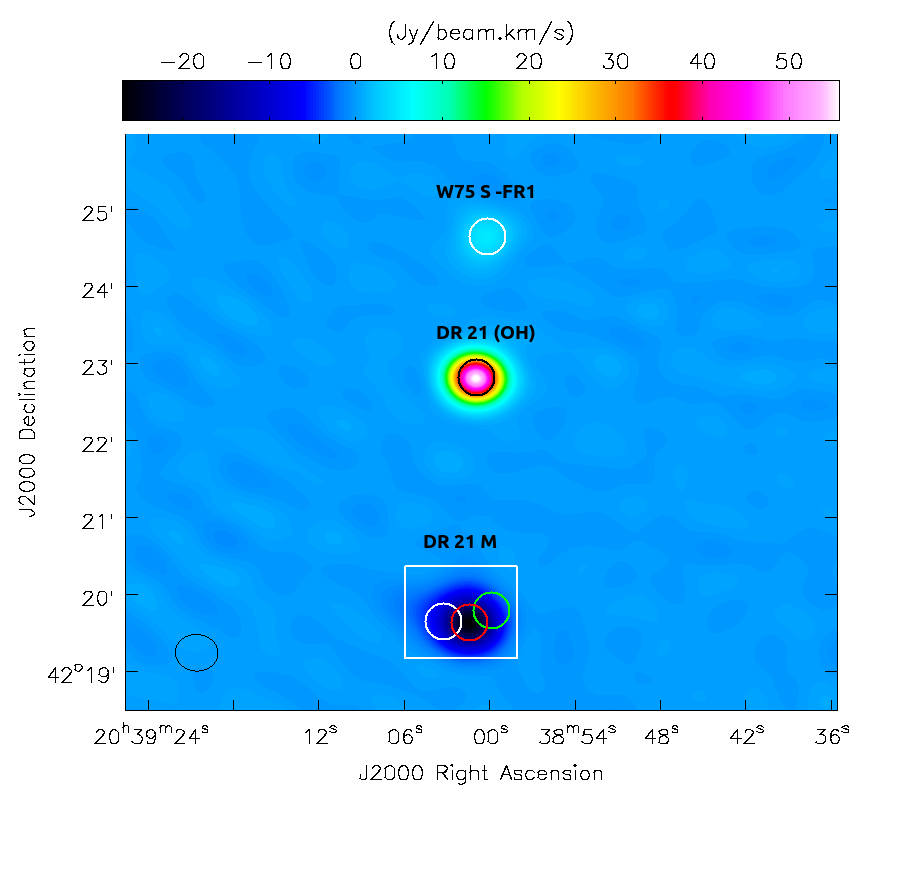}\includegraphics[width=3.6in,height=3.4in,angle=0]{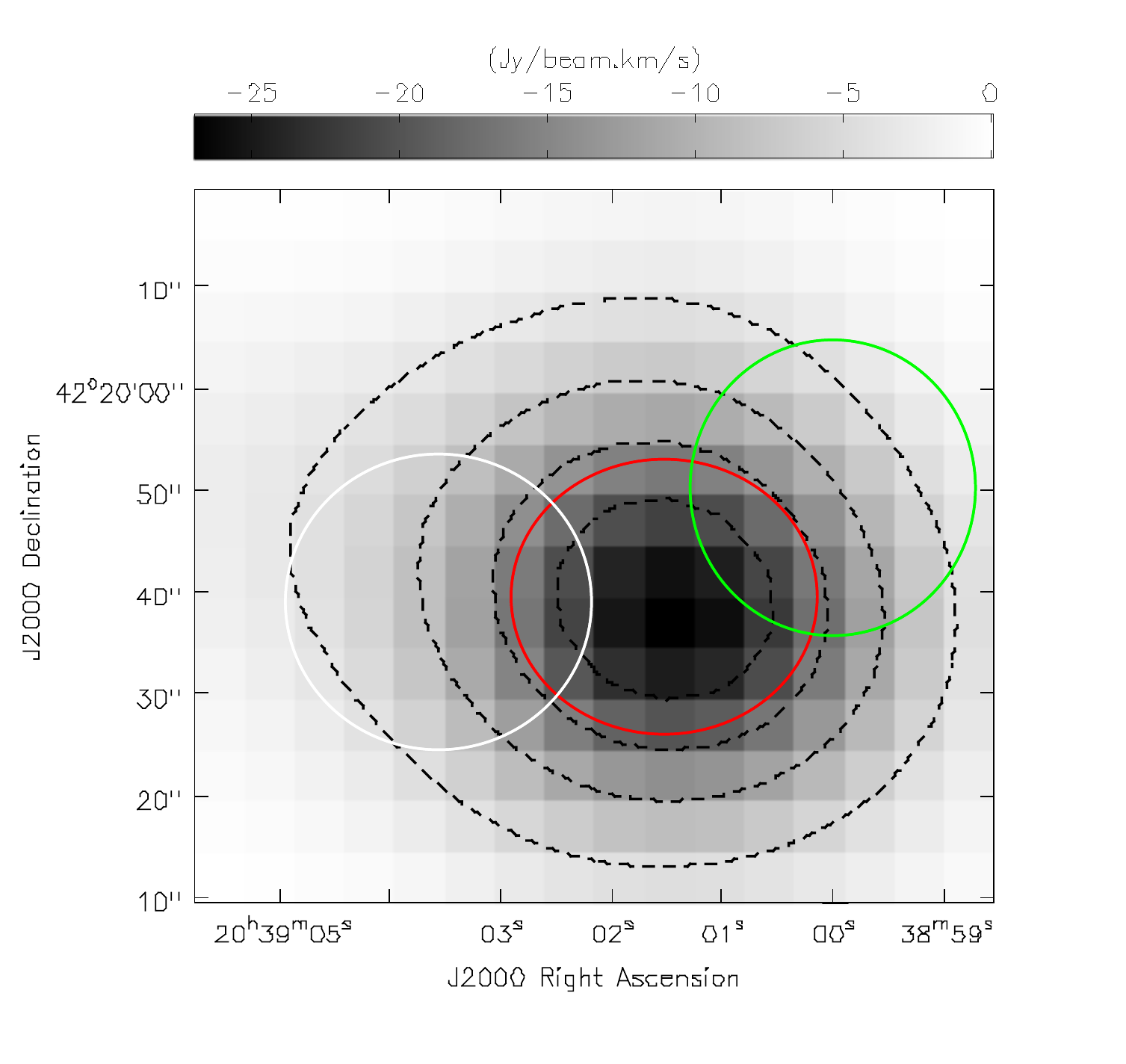}
     \caption{Left: Integrated intensity map [$-86$ to +75~km~s$^{-1}$] of the OH 1665~MHz absorption (toward DR~21~(M); white rectangle) and maser emission (toward DR~21~(OH) and W~75-FR1). The regions toward which maser emission is detected are marked by the black, and white circles which correspond to the positions of DR~21~(OH) ($\alpha = 20{^{\rm h}} 39{^{\rm m}} 01\rlap{.}^{{\rm s}}57, \delta = +42^\circ 22' 45\rlap{.}''74$) and 
      W~75S-FR1 ($\alpha = 20{^{\rm h}} 38{^{\rm m}} 59\rlap{.}^{{\rm s}}99, \delta = +42^\circ 24' 35\rlap{.}''67$), respectively. Right: Zoomed-in view of the DR~21 (M) region showing the marginally resolved continuum emission by dashed black contours (contour levels: 1.24, 2.48, 3.73, and 4.97~ Jy~beam$^{-1}$), along with the integrated absorption in grayscale. The three different colored circles mark the positions from where optical depth spectra are taken from, for comparison. }

    \label{fig:fig3}
\end{figure*}

The primary data for this work are taken from the data archive of the VLA. The observations were carried out over seven observing runs in July and August, 2014 as part of the regular proposal VLA/14A-031. The main observational parameters are summarised in Table~\ref{tab:table1}. For these observations, 3C~286 and 3C~138 were used as flux, delay, and bandpass calibrators, and J2052+3635 was used as the phase calibrator. The phase calibrator was observed for ${\sim 1.2}$ minute for every ${\sim 9.8}$ minute scan of the target source. Initial flagging and calibration are carried out using the scripted CASA (Common Astronomy Software Applications package) pipeline of the National Radio Astronomical Observatory (NRAO)\footnote{The NRAO is operated by Associated Universities Inc., under a collaborative agreement with
the US National Science Foundation.}, with CASA version 4.7.2 and pipeline version 1.3.11, modified suitably for spectral line analysis. We have restricted our analysis to the two spectral line windows centred at the 1665~MHz, and 1667~MHz lines of OH. Flagging, calibration and imaging for the two spectral windows are done entirely independently. For some observations, additional iterative manual flagging of the data and calibration was found to be required. Data from the ``line free'' channels are combined to produce the continuum visibility data. Interactive imaging and deconvolution of the continuum data are done using the task \textsl{\textbf{clean}} to produce the continuum image. Here, uniform weighting (Brigg's robust parameter = $-$2 in \textsl{\textbf{clean}}) of the visibility data is found to produce a marginally better image over the natural weighting. Standard 
self-calibration procedure is also applied to improve the calibration and image quality. Initial calibration, imaging and self-calibration is done separately on data from each day's observations before combining the calibrated data sets, and final continuum imaging. Similarly, continuum subtraction is done in the visibility domain  from each day's data using the CASA task \textsl{\textbf{uvcontsub}}. Final combined image cubes in Stokes I and V are created from the combined data. From 7~hr 50~min on-source time, a noise level of ${\sim 30}$~mJy~beam$^{-1} ({\sim 3\sigma}$) was achieved (for a channel width of 0.98~kHz). Please note that no deconvolution was done for Stokes V, as the signal in each spectral channel is too weak. Finally, we use the CASA task \textsl{\textbf{exportfits}} to convert the image cubes to standard FITS. The resulting map of the integrated 1665~MHz OH line is displayed in Fig.~\ref{fig:fig3}. It shows deep absorption toward DR 21 (M) and maser emission toward DR~21(OH) and W~75-FR1 north of it. 
Further analysis and magnetic field estimation is done using the MIRIAD (Multi-channel Image Reconstruction, Image Analysis, and Display) package \citep[]{1995ASPC...77..433S}.

\subsection{Ionized carbon line data}\label{subsec:CII}
The study of \citet{Ossenkopf2010} presented, a.o., spectra of the $158~\mu$m  ${^2P_{3/2}}-{^2P_{1/2}}$  fine structure line of ionized carbon 
%C{\sc ii} %
C$^{+}$
(rest frequency 1900.547~GHz) observed toward DR 21 with the Heterodyne Instrument for the Far-Infrared (HIFI) aboard the Herschel Space Observatory. For details, see \citet{Ossenkopf2010}. A data cube of spectra mapped around DR~21, kindly provided by V. Ossenkopf, was used to generate the spectrum used in Fig.~\ref{fig:figcii} with an effective resolution of $40^{\prime\prime}$, not too different in size  from that of the restoring beam of our VLA data ($33^{\prime\prime}\times28^{\prime\prime}$).

\subsection{Hydrogen radio recombination line data}\label{subsec:HRRL}
In the framework of a larger scale projects, we observed the H 72$\alpha$ hydrogen recombination line (HRRL) toward DR~21 with the 100-m telescope at Effelsberg, Germany\footnote{The 100-m telescope at Effelsberg is operated by the Max-Planck-Institut f{\"u}r Radioastronomie (MPIfR) on behalf of the Max-Planck Gesellschaft (MPG).} The observations were carried out in position-switching mode with the S20mm receive on 2019 August 23 (project id: 08-19). The S20mm receiver is a double-beam and dual-polarization receiver operating in the 12--18~GHz range that has, at 17.25822~GHz, the frequency of the H 72$\alpha$ RRL, a HPBW of 42\arcsec, similar to that of our VLA and Herschel/HIFI data. The spectrum was analysed with the facility Fourier transform spectrometer (FFTS).
NGC~7027 was used as the absolute flux calibrator, and the flux calibration accuracy is estimated to be within $\sim$10\%.
\section{Results: Magnetic field estimation}\label{sec:results}

Magnetic field in molecular clouds split each of the OH rotational lines into three lines. Now, if the splitting exceeds the line width, and three distinct lines are detectable, then, from the line separations and line ratios for different polarizations, one can estimate the magnetic field strength ($B$) as well as the field orientation ($\theta$) with respect to the line of sight. However, even for narrow maser lines, due to Faraday depolarization in the emission region, all three lines are rarely detectable \citep[see, e.g.,][and references therein]{sault90}. On the other hand, if the magnetic field is weak, the splitting, in general, is smaller than the Doppler width. In this case, from the observed circular polarization, it is possible to measure only the line of sight component of the magnetic field ($B_{\text{LOS}} = B\cos\theta$). Here also, the linearly polarized signal (Stokes Q and U) can, in principle, be used to derive the magnetic field component on the plane of sky (i.e. $B\sin\theta$) as well as the position angle ($\chi$). But, for small splitting, Q and U depends on the second derivative of specific intensity ($I_\nu$) with respect to $\nu$. Thus, the current instrumental sensitivity is not high enough to obtain full three dimensional magnetic field structure from the Zeeman splitting measurements \citep[]{crutcher93, sault90}. There are few complementary methods (e.g. measurement of dust polarization, the ion to neutral line width ratio, or estimating non-thermal broadening) which provide the full information under certain assumptions \citep[see][]{cho16, crutcher12, heiles12, crutcher04, houde00a, houde00b,lp08,ak19,balser16}. Alternatively, one may also use Bayesian analysis for a large number of observations to draw statistical inference \citep[e.g.][]{crutcher10, crutcher99}. 

For the prevalent astrophysical scenario, where the splitting is small compared to the line width, the circular polarization profile (Stokes V spectrum) is proportional to $B_{\text{LOS}}$ and the derivative of the intensity profile (Stokes I spectrum), giving rise to the characteristic ``S'' shaped signature in the Stokes V spectrum (as the unshifted $\pi$ components of the two circular polarizations cancel out identically). In the presence of instrumental leakage, the Stokes V spectrum gets further modified to 
\begin{eqnarray}
V=aI+\frac{b}{2}~\frac{\text{d}I}{\text{d}\nu} \,,
\end{eqnarray}
where the first term is due to polarization leakage, $a$ is a dimensionless constant and $b=zB$cos$\theta$, where $z$ is the Zeeman splitting factor of the transition and $\theta$ is the angle between the line of sight and the magnetic field direction \citep[]{momjian17,troland82}. Note that here we are using the standard convention that a positive value of $B_{\text{LOS}}$ indicates that the field is pointing away from the observer. For positive $B_{\text{LOS}}$, Stokes V and $\frac{\text{d}I}{\text{d}\nu}$ have the same signs, implying that the left circular polarized (LCP) line is shifted to a higher frequency with respect to the right circular polarized (RCP) line. We used the task \textsl{\textbf{zeestat}} in the MIRIAD package to fit the Stokes V profile with the Stokes I profile derivative numerically to estimate the magnetic field. We note that there are several variations in the details of the fitting technique - least-square vs. maximum likelihood method (MLM), ``one-sided'' or ``two-sided'' derivative of the Stokes I profile, `spatial averaging' or `spatial summing' etc., implemented with different heuristic in different data analysis tools. Each of the methods have their own advantages and limitations, some of which are discussed in more detail in \citet{greisen17a,greisen17b,2003ASSL..285..109G}, \citet{killeen92}, and \citet{sault90}. For example, MLM with two-sided derivative is found to be the least biased, except in the case of very narrow lines with sharp edges, where the two-sided method is less accurate. The final values reported here are obtained by fitting the Stokes V spectra from the combined seven day's data, using MLM with two sided derivative and spatial summing option. Additionally, we carefully considered data from each single day, and checked for any artefacts in the profiles. To further cross-check our results, we also repeated the analysis in NRAO AIPS (Astronomical Image Processing System, version 31DEC17) using the AIPS task \textsl{\textbf{ZAMAN}}. The magnetic field values estimated using MIRIAD and AIPS are found to be consistent within the uncertainties. Moreover, the estimation of $B_{\text{LOS}}$ is done completely independently for both the 1665~MHz, and 1667~MHz lines. \citet{brauer17} carried out a radiative transfer analysis of the 1665~MHz and 1667~MHz Zeeman splitting transitions of OH using detailed numerical simulations. These authors concluded that for typical molecular cloud parameters (density, temperature, magnetic field) there is no significant uncertainty due to the analysis method, although for $B\gtrsim300 \, \mu$G it can be about 10\%. In our case, the uncertainty estimated from the fit to the Stokes V profile is larger than 10\%, and is dominated by the sensitivity of the observation.

\begin{figure*}
  \includegraphics[width=3.7in,angle=0]{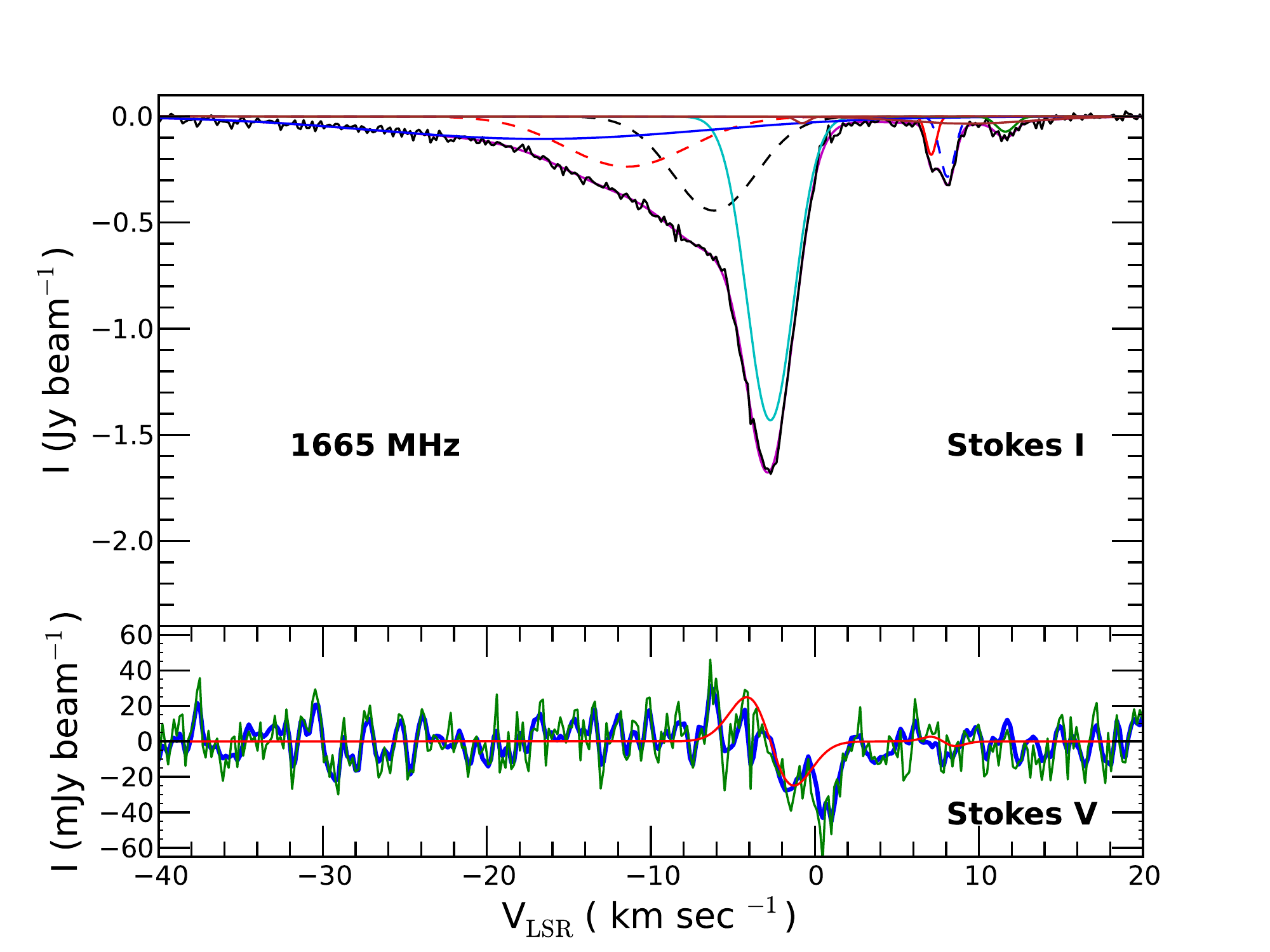}\includegraphics[width=3.7in,angle=0]{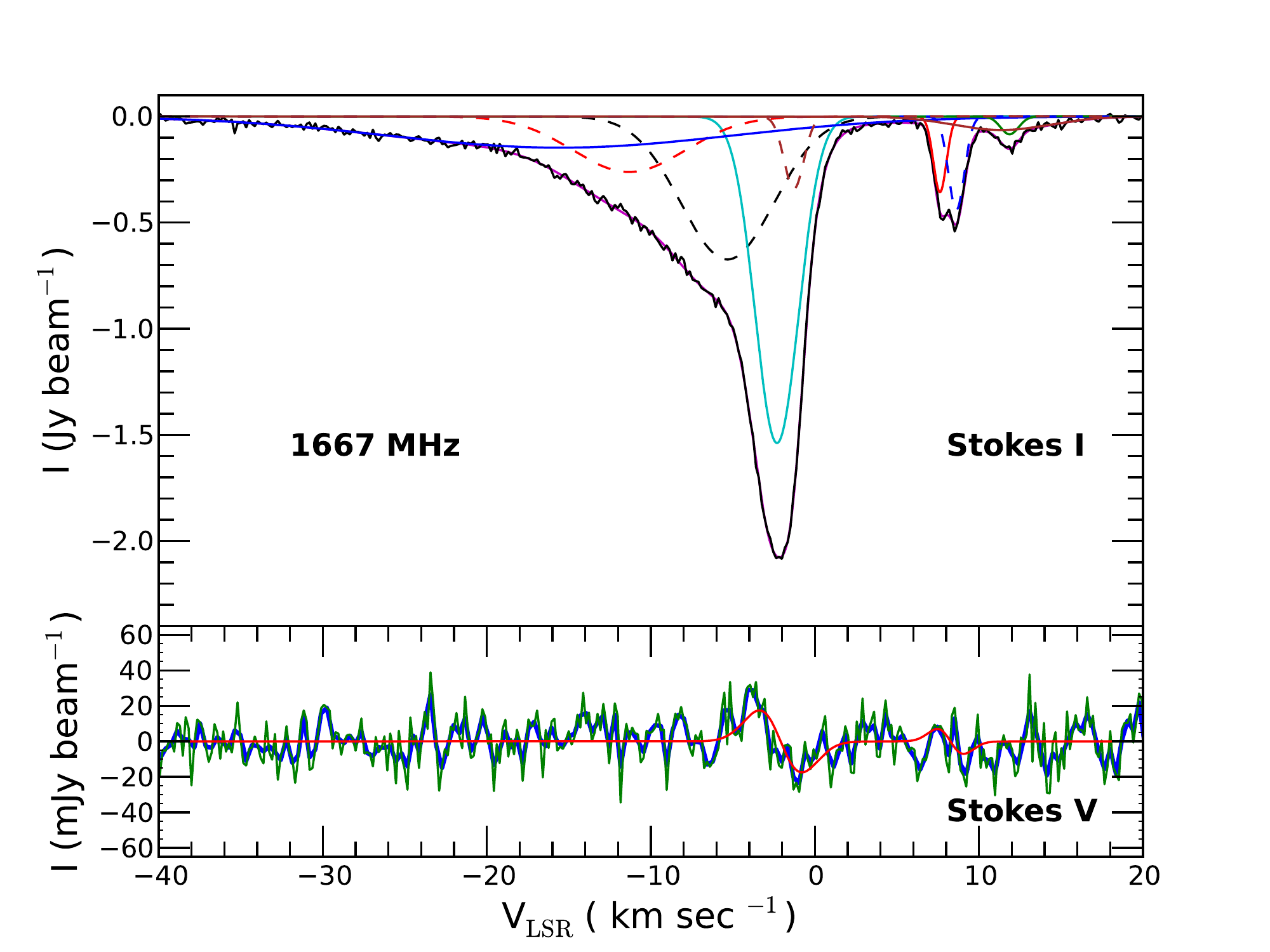}
  \caption{Left: OH 1665~MHz Stokes I (upper panel) and Stokes V (lower panel) profiles toward DR~21~(M). Solid black and maroon lines are the observed spectra and fitted curve, respectively in the upper panel. The nine Gaussian components used to fit the Stokes I profile are also shown. In the bottom panel, the solid green, and blue lines are the observed V profiles without, and with Hanning smoothing, whereas the red line is the fit to the Stokes V profile for the estimated B$_{\text{LOS}} = + 137 \pm 21 \, \mu$G associated with the deep, narrow, negative velocity component (see text for details). Right: Same, but for the OH 1667~MHz profiles, with the best fit B$_{\text{LOS}} = + 123 \pm 24 \, \mu$G.}
  
\label{fig:fig1}
\end{figure*}

In Fig.~\ref{fig:fig1} we display the Stokes I and V profiles from the OH 1665~MHz, and 1667~MHz transitions extracted from the position marked by the green circle, toward DR~21~(M) at 
%($\alpha = 20^h 38^{m} 59.57^{s}, \delta = +42^\circ 19^{\prime} 50.82^{\prime\prime}$) 
$\alpha = 20{^{\rm h}} 38{^{\rm m}} 59\rlap{.}^{{\rm s}}57, \delta = +42^\circ 19'50\rlap{.}''82$, where the Zeeman splitting signal is the strongest. The Stokes I absorption profiles presented in the upper panels are fit using nine Gaussian components (see Sect.~\ref{subsec:ohkinematics} for more details), while the solid red curve in the lower panel represent fits to the Stokes V profiles, following the method described above. The best fit $B_{\text{LOS}}$ values for the deep, narrow component at negative velocities (${v_{\text{LSR}} \sim -2.3}$~km~s$^{-1}$), $+137 \pm 21 \, \mu$G and $+123 \pm 24 \, \mu$G, are consistent within the measurement errors for both the 1665~MHz and 1667~MHz lines, respectively. For the narrow and weak component seen at positive velocities (${v_{\text{LSR}} \sim 8.6}$~km~s$^{-1}$), there is indication of a characteristic ``S'' shaped signal in the Stokes V profile, but the uncertainty in the magnetic field estimated from each of the transitions is larger (see Table~\ref{tab:table34}). 
In addition to the ``S'' shaped signal, we also note the presence of a feature in the Stokes V spectrum of the 1665 MHz transition at ${v_{\text{LSR}} \sim 0 - 2}$~km~s$^{-1}$. This is likely to be due to residuals from the deconvolution and subtraction of the strong emission from the DR~21~(OH) maser site (see Appendix~\ref{app1} for details). This velocity range is not included for estimating the magnetic field strength. Note that this feature is absent in the 1667 MHz Stokes V spectrum. Moreover, the ``S'' shaped signal seen in the 1667~MHz Stokes V spectrum gives magnetic field strength that is consistent with that obtained from the 1665~MHz spectrum, making the measurements reliable.

\begin{figure}
\includegraphics[width=0.5\textwidth]{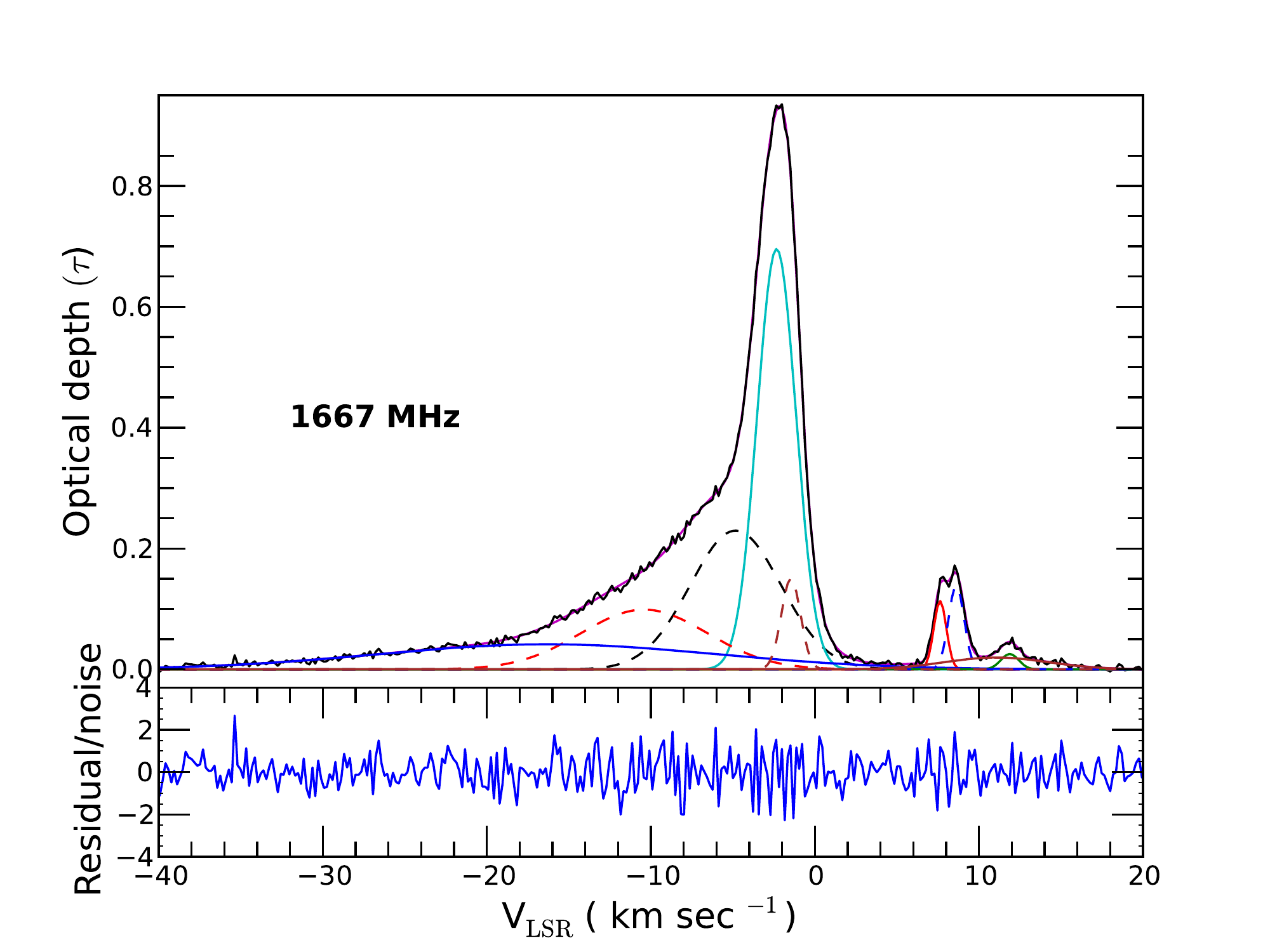}
 \caption{Upper panel: Representative OH 1667~MHz optical depth profile toward the region where magnetic field is significantly detected (green circle in Fig.~\ref{fig:fig3}) in DR~21~(M). The solid black line is the observed spectrum and the maroon line is the best fit model (with nine Gaussian components; details of the fit parameters are in Table~\ref{tab:table2}). Lower Panel: Ratio of the residual to the noise measured from the optical depth spectrum.$^a$}
 
$^a$ Multi-Gaussian fitting are done using python scipy module and figures are made using matplotlib \citep[]{2007CSE.....9...90H}.

\label{fig:fig2}
\end{figure}

\begin{table}    
  \begin{center}

    \caption{Multi-Gaussian fit parameters to the OH 1667~MHz optical depth spectra toward DR~21~(M).}
    \label{tab:table2}
    \begin{tabular}{llccc} 
\hline
Line& $\tau_{\rm peak}$ & $v_{\text{LSR}}$\ & $\Delta v$\,$^a$ & $N_{\rm OH}$\\
type& & (km~s$^{-1}$) & (km~s$^{-1}$) & ($\times 10^{15}$ cm$^{-2}$)\\
\hline
\vspace{1.0 mm}\\
Deep-narrow& $0.69$ & $-2.33$ & $2.78$ & \\
negative& $0.15$ & $-1.46$ & $1.36$ & $12.86$\\
\vspace{1.0 mm}\\
Intermediate& $0.23$ & $-4.83$ & $6.52$ & \\
negative&$0.09$ & $-10.41$ & $9.06$ & $14.47$\\
\vspace{1.0 mm}\\
& $0.03$ & $+11.87$ & $1.27$ & \\
& $0.14$ & $+8.62$ & $1.10$ & $2.59$\\
Positive& $0.02$ & $+11.05$ & $7.41$ &\\
& $0.11$ & $+7.63$ & $0.89$ & \\
\vspace{1.0 mm}\\
Negative-wing& $0.04$ & $-16.32$ & $24.22$ & $6.02$\\ 
\vspace{1.0 mm}
\\
\hline
    \end{tabular}
  \end{center}

   Reduced $\chi_{r}^{2}$ of this fitting is 0.92.\\
  $^a$ Full width at half maximum.\\

\end{table}

We have detected OH maser lines toward previously known maser sites DR~21~(OH) and W~75S-FR1 (see Appendix~\ref{app1} for details). Using these OH maser lines, as well as the absorption lines toward DR~21~(OH), we also critically checked if the ``beam squint'', the relative offset between the primary beam pattern in both polarizations significantly affects the results. The velocity gradient of the spectral line signal across the field along with beam squint may create a false Zeeman-like pattern \citep[e.g.][]{crutcher93}. But this effect is expected to be identical for both the 1665~MHz, and 1667~MHz lines, whereas for real Zeeman pattern, the splitting ratio is $5/3$ for the two lines. We do not see such an identical shift to both the lines, a telltale sign of beam squint effect, in the maser and the absorption spectra. Hence we expect that the signal we observe is due to true Zeeman splitting, and not an artifact due to beam squint.

\section{Discussion}\label{sec:discussion}
\subsection{Gas and dust in DR~21's PDR}\label{subsec:DR21_PDR}

In recent years, with the aid of velocity resolved observations of the fine structure lines of atomic carbon (C~{\sc I} (${}^{3}P_{1}-{}^{3}P_{0}$ and ${}^{3}P_{2}-{}^{3}P_{1}$), [C$^{+}$] and high-$J$ transitions of CO, $^{13}$CO, and HCO$^+$ using HIFI/Herschel the PDR structure of the DR~21 region has been more accurately modeled. \citet{Ossenkopf2010} have modeled this region using the KOSMA-$\tau$ PDR model assuming the DR~21 region to comprise of broadly two ensembles of clumps differing in their gas temperatures, distance from the UV illuminating source and mass fraction. The hotter clumps lie closer to the inner H{\sc ii} region with only a small contribution toward the total mass fraction while the bulk the material is comprised of the cooler clumps. Combined, with the observations, the best fit models result in gas densities between 1.1--1.3$\times10^{6}$~cm$^{-3}$ for the cold and hot components, respectively. This estimate is supported by the H$_2$ densities of $9.1\times10^5$ and $1.9\times10^6$ cm$^{-3}$ that \citet{Motte2007} derive from (sub)mm continuum data for the two clumps associated with the cm emission from DR~21's H{\sc ii} region that are located within the area toward which we image the OH absorption (N46 and N47, respectively, in their nomenclature)

The OH bearing gas is expected to reside in the partially ionized interface region between the central H{\sc ii} region and the general neutral gas, i.e., in a PDR. This is also quite evident from the comparison of the {\it Herschel} HIFI 158~$\mu$m [C{\sc ii}] emission spectrum with the OH 1665~MHz and 1667~MHz absorption spectra. As C$^{+}$ is \textit{the} prime tracer of PDRs, the similarity of the line profiles strongly indicates that the 1665~MHz, and 1667~MHz lines are originating from OH coexisting with C$^{+}$ in the PDR.

\begin{figure}
\includegraphics[width=2.4in,angle=-90]{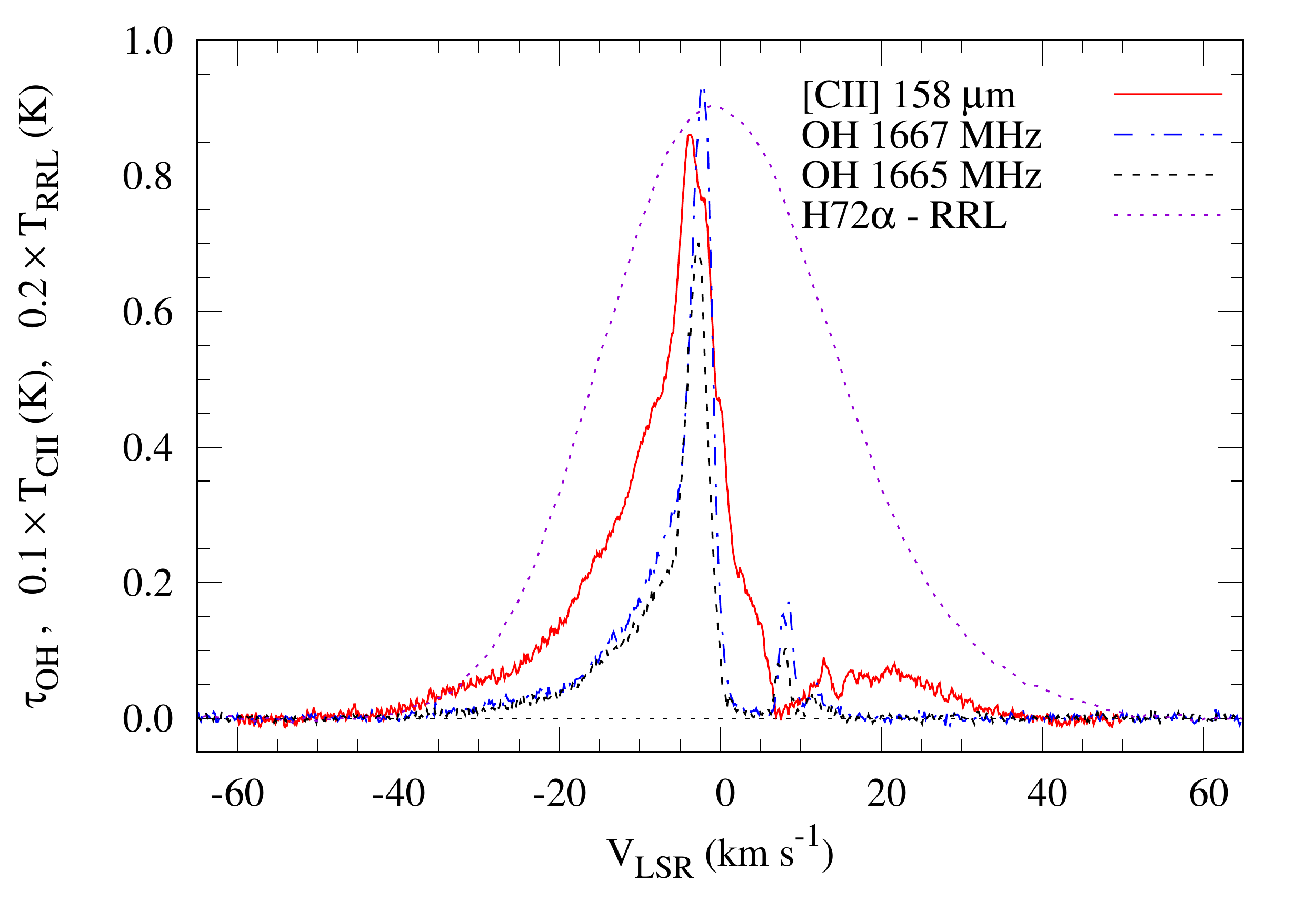}
 \caption{A comparison of the [C{\sc ii}] 158~$\mu$m line profile with that of the OH 1665~MHz, and 1667~MHz optical depth spectra as well as the H 72$\alpha$ line at 17.2582~GHz. The [C{\sc ii}] and HRRL line amplitudes are scaled down by factors of 10 and 20, respectively, for comparison.}
 \label{fig:figcii}
\end{figure}

%{\color{violet}
In Fig.~\ref{fig:figcii}, 
we also show the spectrum of the H 72$\alpha$ RRL. It displays a quite symmetric broad profile covering $-45$ to +50~km~s$^{-1}$, 
that on the low velocity (blue-shifted) side extends to a similar velocity as the OH and the [C{\sc ii}] lines and is centered on a somewhat higher value than the systemic velocity of DR~21's molecular cloud, defined by (sub)mm-wavelength molecular lines, which is between $-3$ and $-1$~km~s$^{-1}$: for example, the 345~GHz CO $J = 3-2$ has a complex multi-peaked spectrum covering $-25$ to $+20$ km~s$^{-1}$, i.e., a narrower range than the HRRL's but centered on the systemic LSR velocity 
\citep{Ossenkopf2010}. 
In contrast, the intensity of the OH absorption and also that of the [C{\sc ii}] emission is very much diminished at (red-shifted) velocities higher than the systemic value.

For OH, this finding has a natural explanation in the fact that the RRL traces \textit{all} the ionizing material of the expanding DR~21 compact H{\sc ii} region, while
on the other hand, OH in absorption  %and [C{\sc ii}] emission %
can naturally only be observed from the part of the also expanding PDR that is moving toward the observer and thus blueshifted. %\textbf{Thus, naturally, only redshifted OH can be observed in absorption. 
The [C{\sc ii}] line also has an asymmetric profile with a marked intensity drop between LSR velocities of $\sim 7$ to $16$ km~s$^{-1}$. Absorption in the lower part of this range corresponds to the extended ``W 75 N 9 km~s$^{-1}$'' cloud discussed in Section \ref{sec:DR21}. We also see absorption in both OH lines in this velocity range. This is expected for both species, as C$^+$ and OH trace PDRs as well as the general ISM. 

It is instructive to compare the [C{\sc ii}] line's profile with the profiles of emission lines from simple molecules. In their Fig. 4, \citet{Schneider2010} present profiles of lines from the low density tracing molecules $^{12}$CO and $^{13}$CO and the higher density tracing HCO$^+$ averaged over the DR~21 filament and ascribe LSR velocity ranges to various kinematic components. Generally, these profiles are centered on DR~21's systemic LSR velocity ($-2.6$~km~s$^{-1}$). In addition, a W~75 N 9~km~s$^{-1}$ cloud component is observed between 5 and 11 km~s$^{-1}$
in \textit{emission} from the low critical density $^{12}$CO and $^{13}$CO $J = 2-1$ lines, while in the HCO$^+$ $1-0$ line an \textit{absorption} dip is observed in that velocity range. This is exactly what one might expect, as the HCO$^+$ line, like the [C{\sc ii}] line (which is also self-absorbed at these velocities), has a significantly higher critical density than the carbon monoxide lines.
% n_crit: CO 2-1 = 2.3 10^3 cm-3, HCO+ 1-0 3 10^4 cm-3 /[CII]: 7 10^3 cm-3 for optically this emission

\subsection{OH and magnetic field in DR~21's PDR\label{subsec:ohabundance}}

For dense PDRs, \citet{hartquist1991} predict gas phase processes to produce substantial abundances of OH (of the order of $10^{-7}$). 
 Via radiative transfer modeling of data for a significant number of OH absorption lines from a variety of rotational states, \citet{jones1994} found the number density of the OH bearing gas, specifically for DR~21 (M), to be $n_{\text{H}_2} = (1.8\pm07)\times10^7$ cm$^{-3}$. These authors derive  a similar value for the OH abundance of $1.6\pm0.7\times10^{-7}$  and 175~K and 125~K for the kinetic and the dust temperatures, respectively. A caveat would be that the collisional rate coefficients employed in this study have undergone several revisions.
In comparison, the gas densities derived by these authors are roughly an order magnitude higher than the values discussed in \S\ref{subsec:DR21_PDR}, which are supported by a variety of data 
and we adopt $10^6$~cm$^{-3}$ as a plausible number in for the H$_2$ density of the OH bearing gas.

Returning to our $B$-field measurement: Ours is the first \textit{direct} measurement of the magnetic field strength in the dense photodissociation region associated with DR~21~(M) 
using OH thermal absorption at 1665~MHz and 1667~MHz and one of the few such measurements in any PDR.  Earlier, \citet{troland16} observing the Orion's veil, a dilute PDR associated with the Orion Nebula and measured magnetic 
fields $\sim 50 - 75 \, \mu$G  in the atomic gas, but as high as ${\sim 350 \, \mu}$G in molecular 
gas using the OH 1665~MHz and 1667~MHz transitions. Additionally, using Zeeman detections of the OH 1665~MHz line toward the central PDR of the M17 H{\sc ii} region/giant molecular complex, 
\citet{Brogan2001} measured the line of sight component of the magnetic field to have absolute values between 150 and 500~$\mu$G.

\begin{figure}
    
    \includegraphics[width=0.475\textwidth]{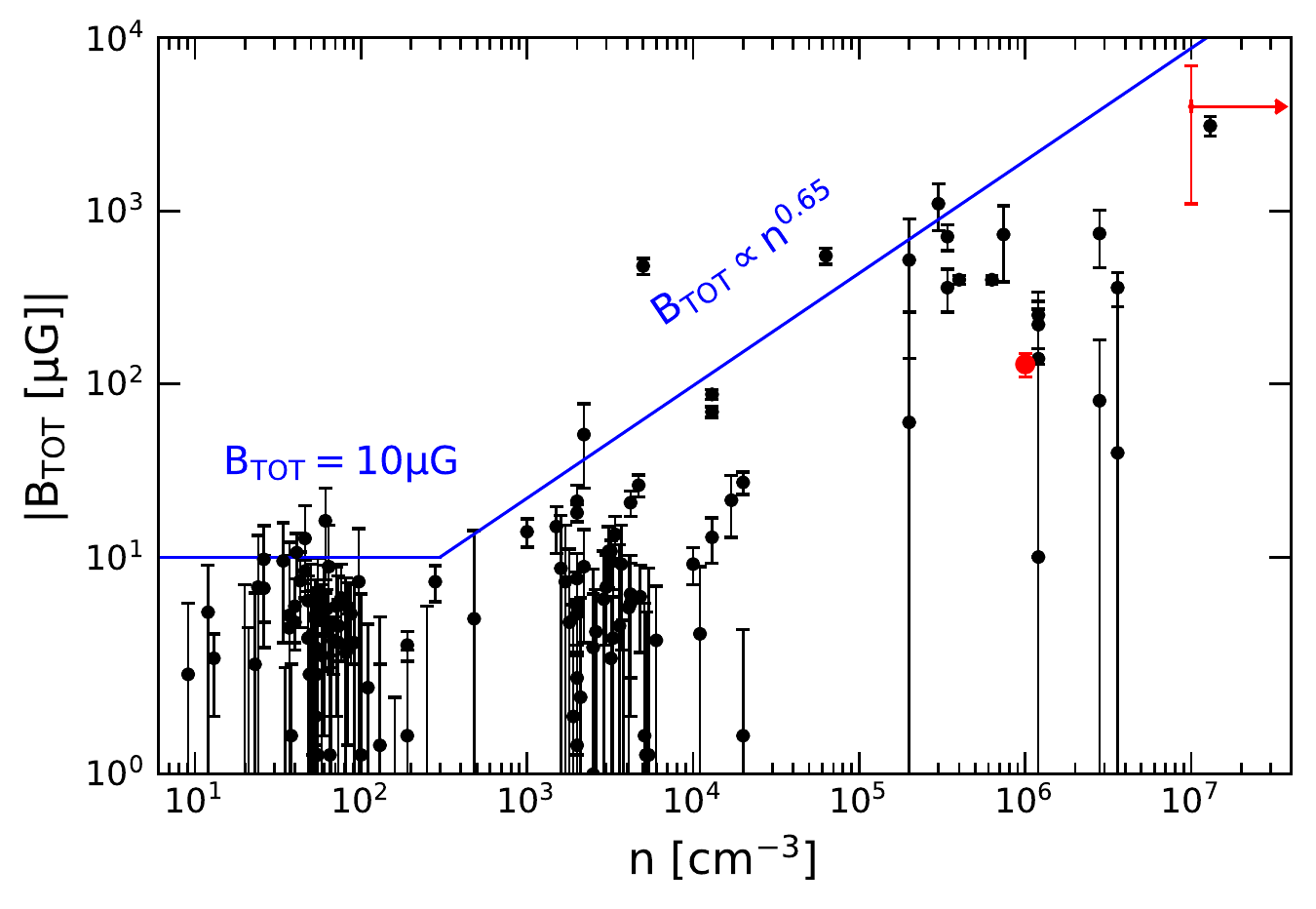}
    \caption{Absolute values of the line of sight magnetic field component, $|B_{\text{LOS}}|$, measured from Zeeman splitting measurements plotted as a function of gas density, $n_{\text H} = n(\text{H{\tiny I}})$ or $2n(\text{H}_{2})$ for H{\tiny I} and molecular clouds, respectively (adapted from \citet{crutcher10}). The black points represent data presented by these authors, %\citet{crutcher10} 
    while the filled red point marks the result for DR~21 discussed in this work. The red bar indicates the entire range of values between 1-8~mG derived by \citet{fish06} for OH maser data and the red arrow reflects the uncertainty of the density of the masing regions. The solid blue line plots the relationship between $|B_{\text{TOT}}|$ and $n_{\text{H}}$ obtained from the Bayesian analysis carried out by \citet{crutcher10}.  }
    \label{fig:Blos_vs_nH}
\end{figure}

\citet{balser16} also reported total field strength {\bf ($B_{\rm TOT}$)} of the order of $100 - 1000 \, \mu$G for PDRs, 
inferred from the non-thermal broadening of the carbon RRLs, an indirect method for $B$-field determinations. It appears to be well established that over many orders of magnitude of the gas density, $n$, the absolute value of the magnetic field strength, {\bf $|B_{\rm TOT}|$}, is proportional to $n^{k}$, where the power-law index, $k$, has a value of 0.65 \citep{crutcher10}. To bring this into context our findings, in Fig.~\ref{fig:Blos_vs_nH} we plot our data point, 0.13~mG, for our assumed H$_2$ density of $10^6$~cm$^{-3}$. As expected, the current measurement, as well as various other entries, falls below the envelope representing the above power law relation.

\citet{Fish2006} analyse and interpret their Very Long Baseline Interferometry (VLBI) data of the 1665~MHz and 1667~MHz OH lines for the 18 high mass star forming regions that were presented in \citet{fish05}. Their conclusions regarding $B$-field strengths are illustrated in Fig. \ref{fig:Blos_vs_nH} and discussed in our Appendix \ref{app1}, in particular for DR~21~(OH). They find a total of 184 Zeeman pairs almost all of which indicate $B$-field strengths between $< 1$ and 8~mG, with 5 outriggers having higher values. Employing the $B$ vs. $n$ relationship described above (but with a power law index of 0.5) \citet{Fish2006}  infer densities corresponding to these $B$-fields, deriving values between $<10^5$ and $10^7$~cm$^{-3}$. In principle, the calculations reported by \citet{cesa1991} and 
\citet{gray1991} allow for inversion over this range of densities, while values at the higher end of this range are more probable. Higher densities would be expected if the masers arose from special, localised high density regions embedded in a lower density OH rich environment. Indeed, for the OH maser emission in 18 high mass star-forming regions they mapped with VLBI, \citet{Fish2006}, report a universal clustering scale for the maser spots  of only $10^{15}$~cm [44 milli arcseconds at DR~21~(OH)]. From the observational side, such scales are difficult to access by observations of thermally excited lines, even in the era of powerful interferometers. Clearly, a detailed excitation/radiative transfer 
study of the physical conditions in OH~bearing environments of (ultra)compact H{\sc ii} regions would be highly desirable that 
would utilise the newly calculated hfs-resolved collision rate coefficients recently presented by 
\citet{Klos2020}.

\subsection{OH column density and excitation}\label{subsec:ohcoldens}

From the 1665~MHz and 1667~MHz absorption lines, the OH column density can be determined using the relation
\begin{eqnarray}
{N_{\rm OH}/T_{\text{ex}}=C \int \tau \,dv \,{\rm cm}^{-2}\, {\rm K}^{-1}}
\label{eqn:eqn_3}
\end{eqnarray}
where $T_{\text{ex}}$ is the excitation temperature (in K) of the respective OH transitions, $ dv $ is in km~s$^{-1}$, and the value of the constant $C$ is $4.1063\times 10^{14}$ and $2.2785\times 10^{14}$~cm$^{-2}$~(km~s$^{-1}$)$^{-1}$~K$^{-1}$ for the OH 1665~MHz and 1667~MHz lines, respectively \citep[]{sarma00,roberts95}. 
In principle, a joint absorption-emission study could provide the excitation temperature. As we have no emission data for DR~21, we resort to the literature for information on $T_{\text{ex}}$. Strong maser emission in the 1665/1667~MHz OH lines is an archetypal signpost for \textit{ultra}compact H{\sc ii} regions, with W3(OH) the first source identified as such \citep{mezger1967}. In addition to the ground-state lines, many hfs lines from rotationally excited levels also are masing, while others show (enhanced) absorption \citep{cesa1991}. In the material around more evolved compact H{\sc ii} regions such as DR~21, the physical conditions no longer support maser action and absorption prevails. Radiative transfer model calculations using such multi-transition observational data as constraints, such as those of
\citet[][see \S \ref{subsec:ohabundance}] {jones1994},
can be used to constrain the physical conditions in OH bearing regions. Unfortunately, that study gives no information of $T_{\text{ex}}$ values predicted by their models. In contrast, the extensive study of \citet{cesa1991} does present, in its Table~2, $T_{\text{ex}}$ values for a fixed $n_{\text{H}_2} = 10^7$~cm$^3$ (close to the value we derived above for DR~21) and a wide range of OH column densities from $5\times10^{13}$~cm$^{-3}$ to $1.6\times10^{17}$~cm$^{-3}$. 
Except for the lowest and the highest values, they find $T_{\text{ex}}$ to be negative, i.e., inversion, for the 1665 and 1667 MHz lines. For these lines, they derive $T_{\text{ex}}$ = 7.7 and 32.3 K, values that are diametrically opposite from those found in an earlier study by
\citet{Guilloteau1985},
32.5 and 8.7, respectively. Given the obvious uncertainties of such estimates, in our case we assume a fiducial value of $T_{\text{ex}} = 25$~K.
to calculate the column densities reported for the different velocity  components in Table~\ref{tab:table2}. This yields column densities in the range of ${\sim 10^{15}}$--$10^{16}$~cm$^{-2}$, lower than the value \citet{cesa1991} quoted above. This is hardly surprising given all the uncertainties and the fact that local thermodynamic equilibrium (LTE), which is a premise for the applicability of Eq.~\ref{eqn:eqn_3}, is clearly not a valid concept for OH in the regions in question.

\begin{table}    
  \begin{center}

    \caption{Estimated line of sight component of the magnetic field in DR~21~(M) from the OH 1665~MHz and 1667~MHz absorption lines.}
    \label{tab:table34}
    \begin{tabular}{lcccc} 
\hline
             &\multicolumn{2}{c}{Deep-narrow\,$^a$} & \multicolumn{2}{c}{Positive\,$^a$}\\
\hline
Line (MHz)   & 1665 & 1667 & 1665 & 1667 \\
Amp. (Jy/beam)& -1.40&-1.54&-0.28&-0.45\\
$B_{\text{LOS}}$ ($\mu$G)  & $137\pm21$&$123\pm24$&$20\pm45$&$104\pm43$\\
 & \multicolumn{2}{c}{$~~130\pm16 ~\mu$G\,$^b$}& \\ %\multicolumn{2}{c}{$~~62\pm31 ~\mu{\rm G}$\,$^b$}\\
\hline
    \end{tabular}
  \end{center}
  $^a$ Component as in Table~\ref{tab:table2}\\
  $^b$ $B_{\text{LOS}}$ values and uncertainties combining both the measurements
\end{table}

\subsection{OH kinematics}
\label{subsec:ohkinematics}

Earlier \citet{cyganowski03} and \citet{immer14} carried out comprehensive studies of the DR~21~(M) region using radio continuum, molecular tracers as well as radio recombination lines, with detailed discussion on champagne flow versus bow shock model for the H{\sc ii} region continuum. General radiative transfer for OH absorption in presence of the Galactic isotropic and discrete source background is discussed in \citet{goss68}. Typically, when the discrete source is the dominant background, the radiative transfer equation \citep[]{roberts97} becomes
\begin{eqnarray}
\label{equn: equn 2}
1+\frac{T_\text{b}(v)}{T_\text{c}} = e^{-\tau(v)} 
\end{eqnarray}
/where $T_\text{b}(v)$ and $T_\text{c}$ are the continuum subtracted brightness temperature of the absorption line and the continuum temperature, respectively. The CASA task \textsl{\textbf{immath}} is used to convert the spectral line data in terms of the optical depth, following Eq~\ref{equn: equn 2}. As an example the OH 1667~MHz optical depth profile is shown in Fig.~\ref{fig:fig2} toward the region where the magnetic field is detected. Nine prominent components, namely two narrow and two moderately broad components at negative velocities, three narrow components and one moderately broad components at positive velocities, and a broad negative velocity wing are evident from the multi-Gaussian modeling of the optical depth profile of the OH spectra as shown in Fig.~\ref{fig:fig2}. The details of the multi-Gaussian fit parameters are summarised in Table~\ref{tab:table2}. There is some variation of the optical depth depth across the background continuum region as shown in Appendix~\ref{app2}, however due to the coarser angular resolution of our observations, we have not been able to study if there is any clear systematic variation of the central velocities of these different components. 

Previous high resolution ($5^{\prime\prime}$) H{\sc i} and CO observations attributed this negative velocity wing to an outflow component emerging from the embedded protostar \citep[e.g.][]{roberts97,fischer80} with variation of central velocity along the outflow axis \citep[NE-SW on the plane of the sky][]{garden92}. \citet{garden91} estimated the mass of the outflow to be ${> 3000~}$M$_\odot$ and energy ${> 2 \times 10^{48}}$~ergs, one of the most energetic outflows observed in such systems. \citet{garden90} and \citet{garden92} have also reported hot ammonia and H$_{2}$O maser near the origin of the outflow (an infrared star I6), and HCO$^{+}$ as well as shocked 2.2~$\mu$m H$_{2}$ emission along the outflow axis.

\subsection{Magnetic field and energetics}\label{subsec:energetic}

While estimating the magnetic field from the Zeeman splitting signal following the method described earlier, the field strength is kept as a free parameter for each of the nine prominent components. Indeed, the possibility of detecting or constraining the field strength is higher for the strong and narrow line. As expected, we did not find any viable solution for the weaker or relatively broader components. As shown in Table~\ref{tab:table34}, we detect the Zeeman signal (magnetic field value with more than 3$\sigma$ significance) in both the 1665~MHz and 1667~MHz but only for the strongest and narrowest component, with an estimated value of $B_{\text{LOS}} = 130\pm16 \, \mu$G from combining both the measurements. For the other narrow component at positive velocities, we only get an upper limit for the magnetic field $B_{\text{LOS}} \lesssim 130 \, \mu$G ($3\sigma$). Previously, \citet{fish06} have determined a $3\sigma$ upper limit of $300 \, \mu$G in the magnetic field strength in DR~21 M's PDR, based on their observations of absorption in the rotationally excited 6030 MHz transition, consistent with the current detection. Other magnetic field determinations toward DR~21 are at variance with those measurements \citep{roberts97} %, on the other hand, 

Based on the measured magnetic field value, one may infer the importance of magnetic field by comparing the magnetic field energy density with the thermal and hydrodynamic energy density. For thermal energy density ($u_{\text{th}} = 3/2 P_{\text{th}} = 3/2 nk_{\text{B}}T$), kinetic temperature is assumed to be ${\sim T_{\text{ex}}}$, and typical $n$(H$_{2}$) ${\sim 10^{6}}$ cm$^{-3}$ in the this PDR region 
%\citep[]{crutcher04}; 
this results in an energy density of $u_{\text{th}} = 5.17 \times 10^{-9}$ dyn~cm$^{-2}$ ($3.23 \times 10^{3}$~eV~cm$^{-3}$). For simplicity, we have assumed the velocity dispersion to be isotropic, and the hydrodynamic energy ($u_{\text{hydro}} = 1/2 \rho v^2$) is estimated to be $\sim 9 \times 10^{-8}$~dyn~cm$^{-2}$ ($\sim 5.7 \times 10^{4}$~eV~cm $^{-3}$). The magnetic field energy density ($u_\text{B} = B^2/8\pi$), on the other hand, is $0.67$, $2.69$ and $22.3\times 10^{-9}$ dyn~cm$^{-2}$ ($0.42$, $1.68$ and $13.9$ keV~cm$^{-3}$) for a magnetic field orientation of
$\theta$ = $0^{\circ}$, $60^{\circ}$ and $80^{\circ}$ 
with respect to the line of sight, respectively. Note that the various energy densities are computed only for the narrow line component for which the magnetic field value has been measured from the OH Zeeman splitting as in Table ~\ref{tab:table2}. For the $\theta$ = $0^{\circ}$, $60^\circ$, $80^\circ$ scenarios, the ratios of thermal to magnetic pressure (similar to ``plasma $\beta$'' parameter in highly ionized medium) are $\sim$ 5.1, 1.3, 0.2 respectively. %From this result, it is clear that the thermal pressure dominates over the magnetic pressure. 

Another important parameter to consider is the mass to flux ratio based on our estimated magnetic field value. The mass to flux ratio over its critical value \citep{nakano78} is given by 
\begin{eqnarray}
{\lambda_c \equiv \frac{(M/\Phi)_{\rm obs}}{(M/\Phi)_{\rm critical}} = 7.6 \times 10^{-21} {\frac{N(\text{H}_{2})}{B_{\text{total}}}}}
\end{eqnarray}
where $N(\text{H}_{2})$ is in cm$^{-2}$, $B_{\text{total}}$ is in $\mu$G, and $(\frac{M}{\Phi})_{\rm critical} = 1/\sqrt{4\pi^2G}$ (also see \citealt{tomisaka98} and \citealt{krumholz11} for a discussion on critical mass to flux ratio).
Using an OH abundance of $\sim 10^{-7}$ \citep{jones1994}, we derive a H$_{2}$ column density for the narrow component of $\sim 1.1 \times 10^{23}$ cm$^{-2}$. Thus, the  normalized mass to flux ratio values are $6.6$, $3.4$ and $1.11$ for $\theta = 0^\circ$, $60^\circ$ and $80^\circ$, respectively. Clearly, with these reasonable assumptions, {\bf $\lambda_c> 1$}, for all possible orientations. So, from these result we can clearly say that we are probing a magnetically supercritical envelope region, where the magnetic field is not dynamically important. This is consistent with magnetohydrodynamic simulations that show that while magnetic field might suppress small-scale fragmentation, it may have little influence on the expansion of the H{\sc ii} region itself (e.g. \citealt{arthur2011}).

Despite these attainable calculations showing the relative importance of magnetic fields in molecular clouds, many more such observations and theoretical studies \citep[see, e.g.][]{mouschovias09,tritsis15} are necessary to address and resolve the questions at hand. In addition, further deeper and higher resolution observations of the DR~21 region will be useful in order to detect (or put tight constraints) the magnetic fields associated with the other components, as well as to carry out spatially resolved estimations of the magnetic field. This will allow us to better understand the field geometry as well as probe the density dependence of the magnetic field in different parts, and thus test various theoretical predictions.

\section{Conclusions}\label{sec:conclusions}

We have presented data on the 1665~MHz and 1667~MHz transitions of OH, in absorption toward DR~21~(M) and showing maser emission in the neighbouring DR~21~(OH) and W~75S-FR1 regions. Our observations and analysis reveal the following:
\begin{enumerate}
%\vspace{0.2cm}
\item We determine a line of sight magnetic field strength of $130\pm16~ \mu$G toward the PDR associated with the DR~21~(M) region.%, which is associated with the strong narrow OH absorption component.\\
\item Out observations of OH absorption, a hydrogen RRL and the [C{\sc ii}] $158~\mu$m line (the quintessential tracer of PDRs) all are consistent with the scenario that OH is co-existing with C$^{+}$ in the PDR, whereas the RRL traces the ionized gas in the compact, expanding H{\sc ii} region. The asymmetry in the [C{\sc ii}] line is due to absorption from the W~75 N 9 km~s$^{-1}$ component.%\\
\item In addition to the narrow and moderate components, 
we detect one negative velocity wing (FWHM $\sim 24$~km~s$^{-1}$), possibly associated with the prominent DR~21 bipolar outflow.
%\vspace{0.2cm}
\item The estimated OH column densities for the various components are in the range from $\sim 10^{15} - 10^{16}$ cm$^{-2}$ toward the PDR surrounding the compact H{\sc ii} region of DR~21~(M), and a total column density of $\sim 3.6\times10^{16}(T_{ex}/25 {\rm K})~{\rm cm}^{-2}$ with significant uncertainty due to the lack of any tight constraint on $T_{ex}$.%\\
%\vspace{0.2cm}

%\vspace{0.2cm}

\item We also detect both 1665~MHz and 1667~MHz OH masers toward DR~21~(OH), and only 1665~MHz masers for W~75S-FR1.%\\ % (see Appendix~\ref{app1}).
%\vspace{0.2cm}

\item We calculate energy densities and mass to flux ratio for DR~21~(M) under some reasonable assumptions, and show that the thermal, hydrodynamical and gravitational energy density dominate over magnetic energy density.
%Also, the magnetic energy alone is insufficient to support the molecular material against self-gravity.
\end{enumerate}
\section*{Acknowledgements}

 We thank the anonymous referee for suggestions and comments that have
helped to improve this work. We would like to thank Volker Ossenkopf for making his C{\sc ii} data available to us. A.K. would like to thank DST-INSPIRE (IF160553) for a fellowship. N.R. acknowledges support from MPG through Max-Planck India partner group grant.  AMJ is a member of the International Max Planck Research School (IMPRS) for Astronomy and Astrophysics at the Universities of Bonn and Cologne.TGSP gratefully acknowledges support by the National Science Foundation under grant No. AST-2009842.
We are thankful to Yan Gong for his help with performing the Effelsberg observations.\\
 
\section*{Data Availability}
The data products from this study will be shared on reasonable request to the corresponding author.

\bibliographystyle{mnras}
\bibliography{dr21oh}

\appendix 

\section{DR~21~(OH) maser emission}
\label{app1}
\begin{figure*}
  \includegraphics[width=3.7in,angle=0]{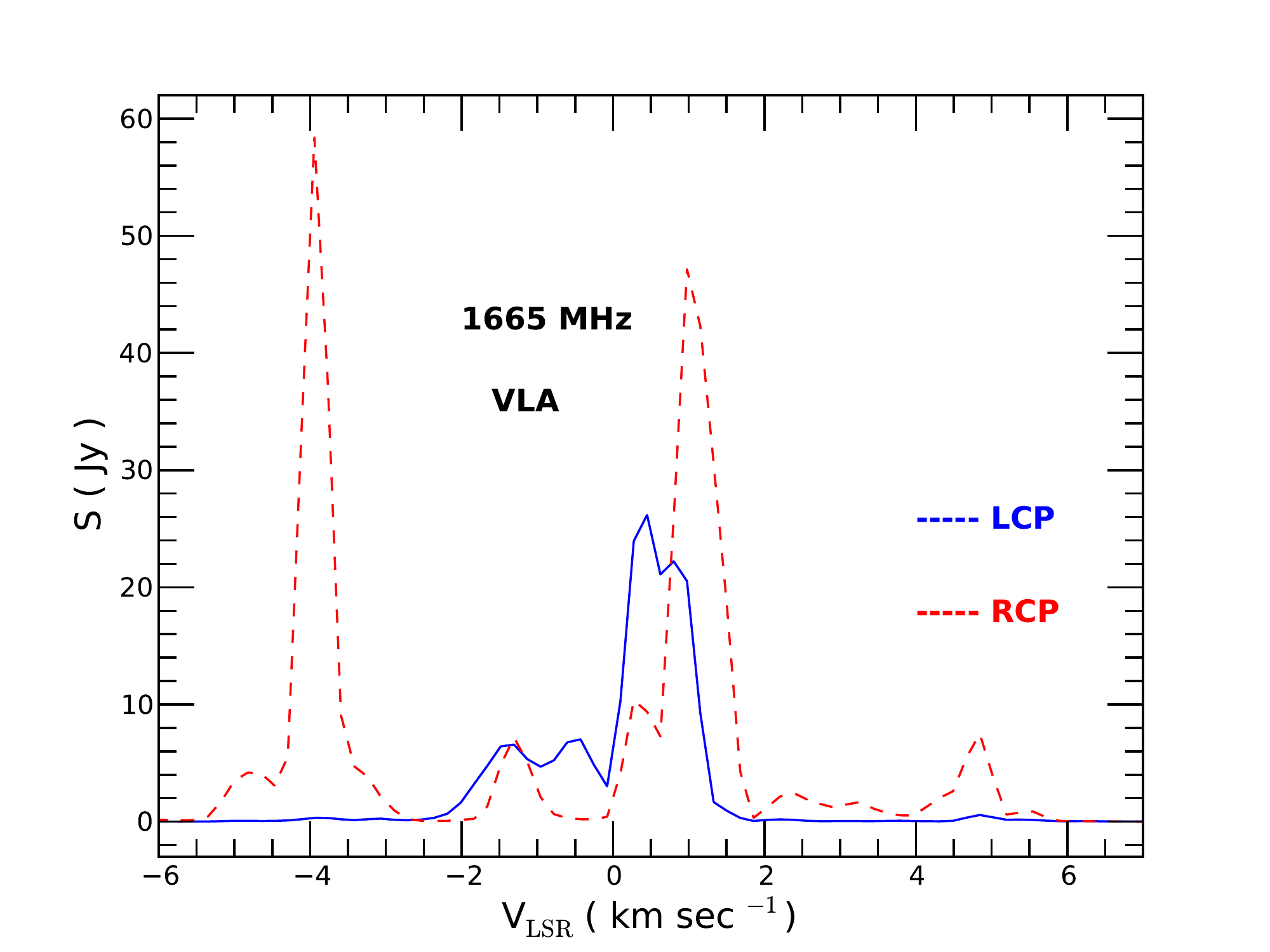}\includegraphics[width=3.7in,angle=0]{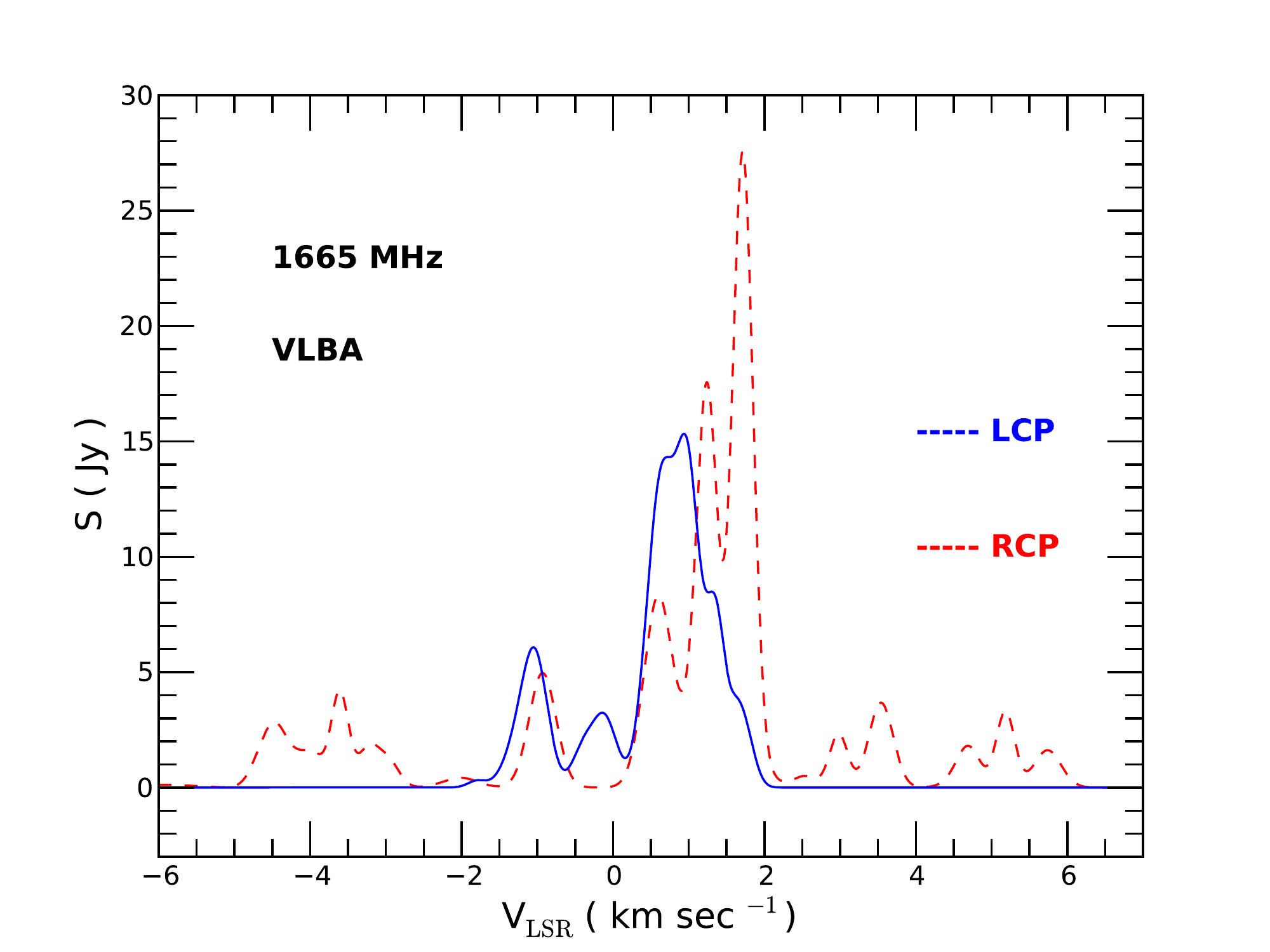}
  \caption{Left: OH 1665~MHz maser lines at the position of DR~21~(OH) from the present VLA data. The Stokes left circular polarization (LCP) and right circular polarization (RCP) components are represented by the solid blue and dashed red lines, respectively. 
 Right: Equivalent spectra observed using the VLBA at an earlier epoch, reconstructed from parameters given in \citet{fish05}. Both sets of spectra show a good correspondence between the distribution of velocity components but show variability in their amplitude.}
\label{fig:fig4}
\end{figure*}

\begin{figure*}
  \includegraphics[width=3.7in,angle=0]{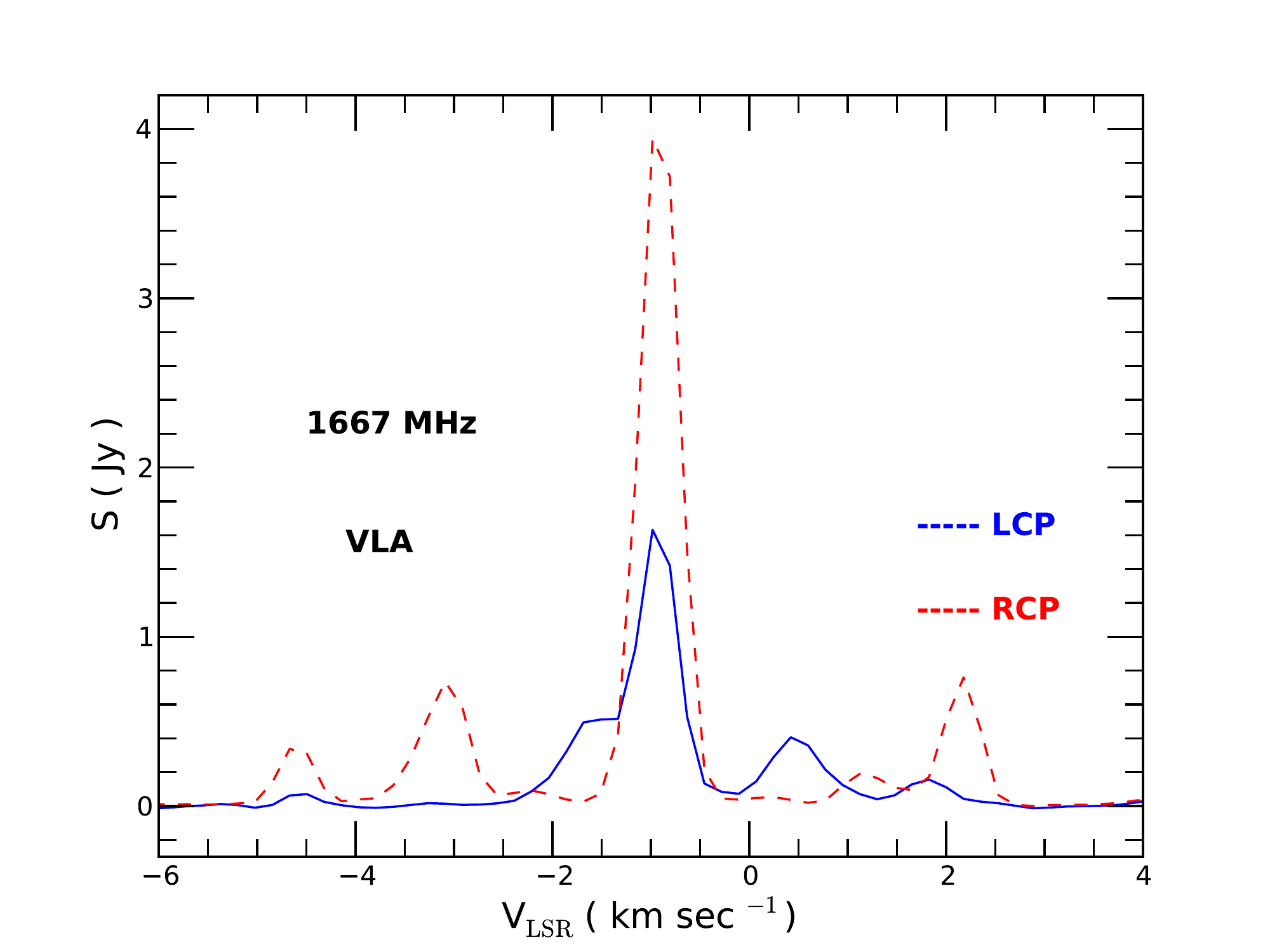}\includegraphics[width=3.7in,angle=0]{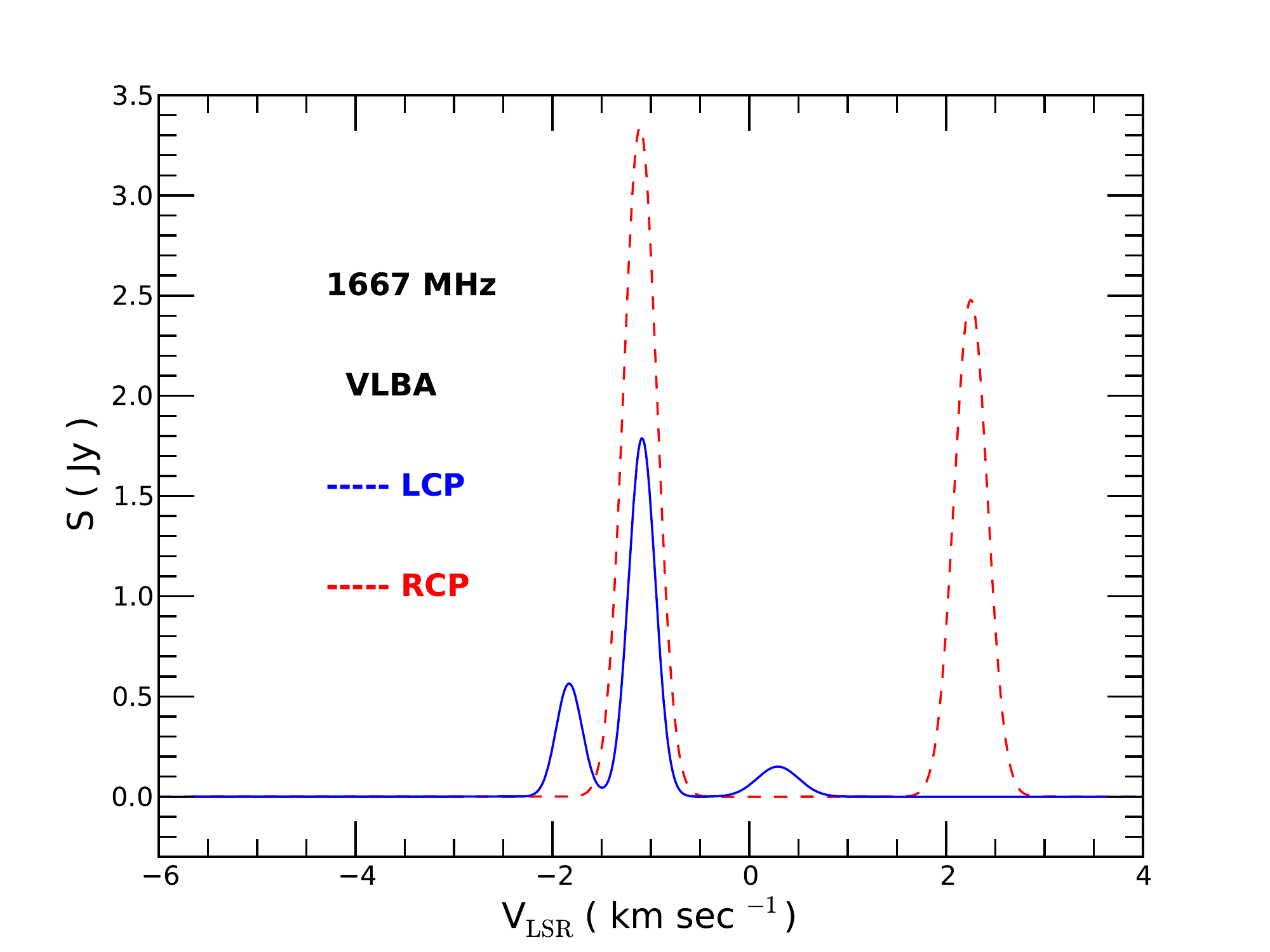}
  \caption{Same as Fig.~\ref{fig:fig4} but for the OH 1667~MHz line.}
\label{fig:fig5}
\end{figure*}
\begin{figure}
\includegraphics[width=3.8in]{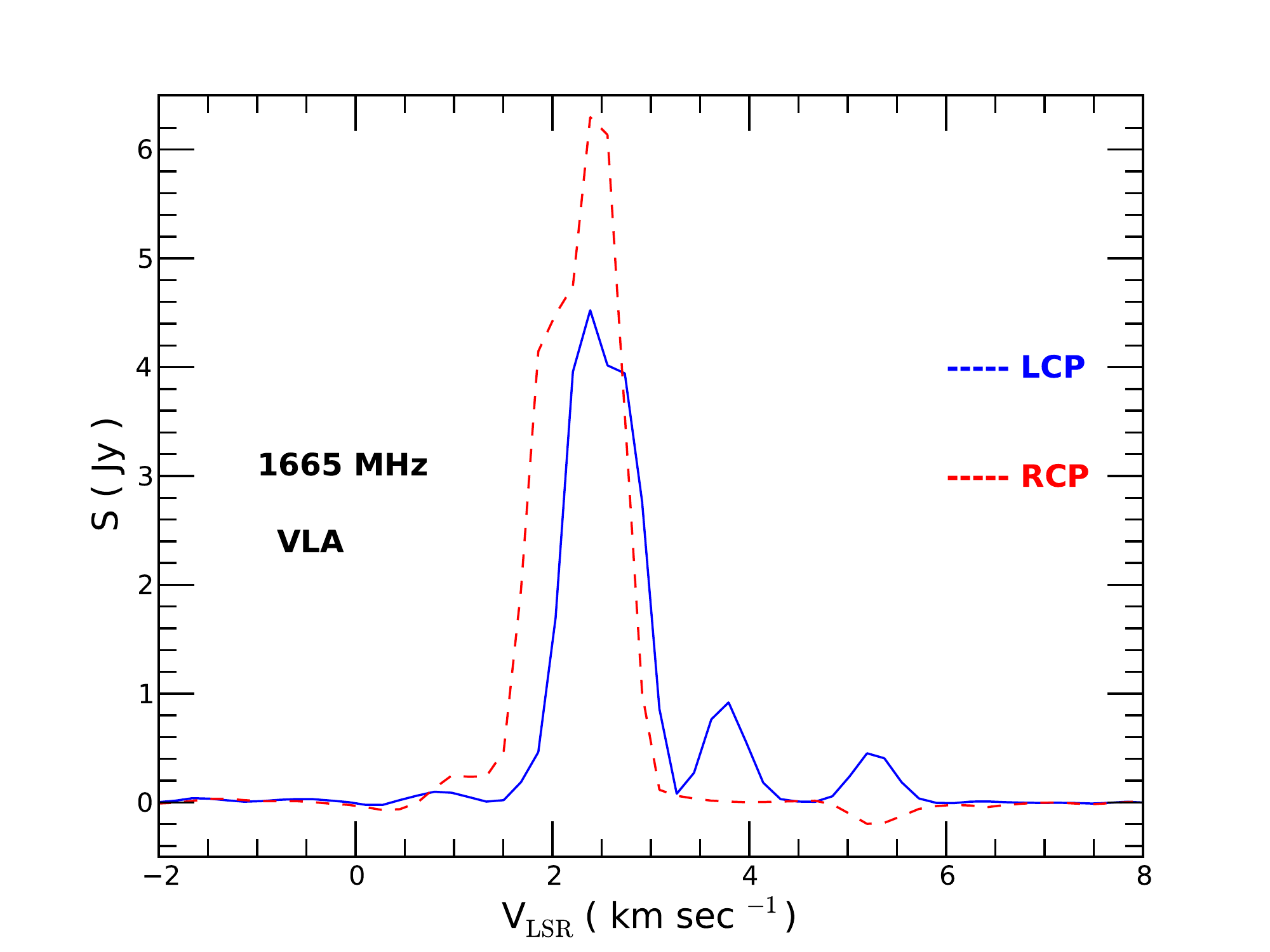}
  \caption{OH 1665~MHz maser spectra at the position of W~75S-FR1 in right and left circular polarization. The corresponding 1667~MHz transition is not detected at this location.}
  \label{fig:fig6}  
\end{figure}

In addition to the OH absorption lines toward DR~21~(M), we also detect the OH 1665~MHz and 1667~MHz maser lines toward previously known maser sites DR~21~(OH) and W~75S-FR1 which were also covered by our imaging (see Fig.~\ref{fig:fig3}). For the W~75S-FR1 region, only the 1665~MHz maser line is detected. The observed spectra are shown in Figs.~\ref{fig:fig4}--\ref{fig:fig6}. These narrow, bright and variable maser lines originate in high density compact regions. As mentioned earlier, full spectro-polarimetric VLBI observations of these lines are required to measure the magnetic fields as well as the velocity structures and dynamical evolution of these small ($\sim 100$~au sized) compact regions. This, however, necessitates high spatial resolution observations to disentangle emission from multiple maser spots in the same region at slightly different velocities. 
As even its name says, DR 21(OH) is a classic OH maser region. Relatively recent studies of the 1665 and 1667~MHz OH hfs lines by \citet{argon00,fish03,fish05} with the VLA and the Very Long Baseline Array (VLBA) show a velocity range of the emission comparable to that observed by us (see further below).
The Zeeman splitting factors quoted in the \S\ref{sec:intro} correspond to 0.59 and for the 1665 and 1667 MHz lines, respectively. 
With the few mG fields in maser regions, this causes differences of the RCL and LCP signals of the order of a few km~s$^{-1}$. As first pointed out by \citet{Cook1966}, given the velocity structure and turbulence in maser clumps, this can affect the gain paths for the RCP and LCP signals differently, which may result in very different intensities.
This make $B$-field determinations based from single dish measurements of the 18~cm OH lines very uncertain or impossible. On the other hand, very reliable field measurements are possible if VLBI data are available, which yield sub-milliarcsecond accuracy for maser spot positions and RCP to LCP registration accuracy \citep{fish05}. A field can be determined if RCP and LCP components appear at the same position in the sky. For DR 21(OH) \citet{fish05} find three clusters of maser spots separated by $\si 1.5^{\prime\prime}$ (2200~au) from each other, each containing several components for some of which they determine $B$-field strengths. They find values of $-5.3$ to $-3.8$~mG, $-7.6$~mG and +5.6 to +6.6~mG, respectively, indicating a field reversal. We include the above mentioned field strengths in Fig.~\ref{fig:Blos_vs_nH} and our discussion in \S\ref{subsec:ohabundance}.
We note that these field strengths are an order of magnitude
higher than the values determined for two dense molecular cores in DR~21~(OH) discussed in\S\ref{sec:DR21}, again indicating significantly enhanced densities in maser emitting clumps.

Of course, the data presented here, observed with the VLA in D configuration, does not offer adequate angular resolution to separate the maser spots. As a consistency check, we compare the observed spectra resulting from our observations with the reconstructed VLBA spectra based on published parameters taken from \citet{fish05}. While this serves as a qualitative check for consistency, %we note here that, due to very different spatial resolution, 
a one to one comparison of the velocity components does not make sense. In spite of that, Fig.~\ref{fig:fig4} shows very good agreement of the central $v_{\rm LSR}$ of different components (in both polarizations), and significant variation of relative amplitudes for the 1665~MHz maser lines. For the 1667~MHz lines also, as shown in Fig.~\ref{fig:fig5}, we also see variability (e.g., a much lower intensity of the prominent $2.25$~km~s$^{-1}$ RCP component in our VLA spectrum; a few of our components are not reported in \citet{fish05}.% Further discussion 
\begin{comment}
Due to the compactness of the emitting regions, a kinematical study with maser lines would require much higher spatial resolution than that of this study.
\end{comment}
\section{OH Optical depth spectra in different regions}
\label{app2}

\begin{figure}
  \includegraphics[width=3.7in,angle=0]{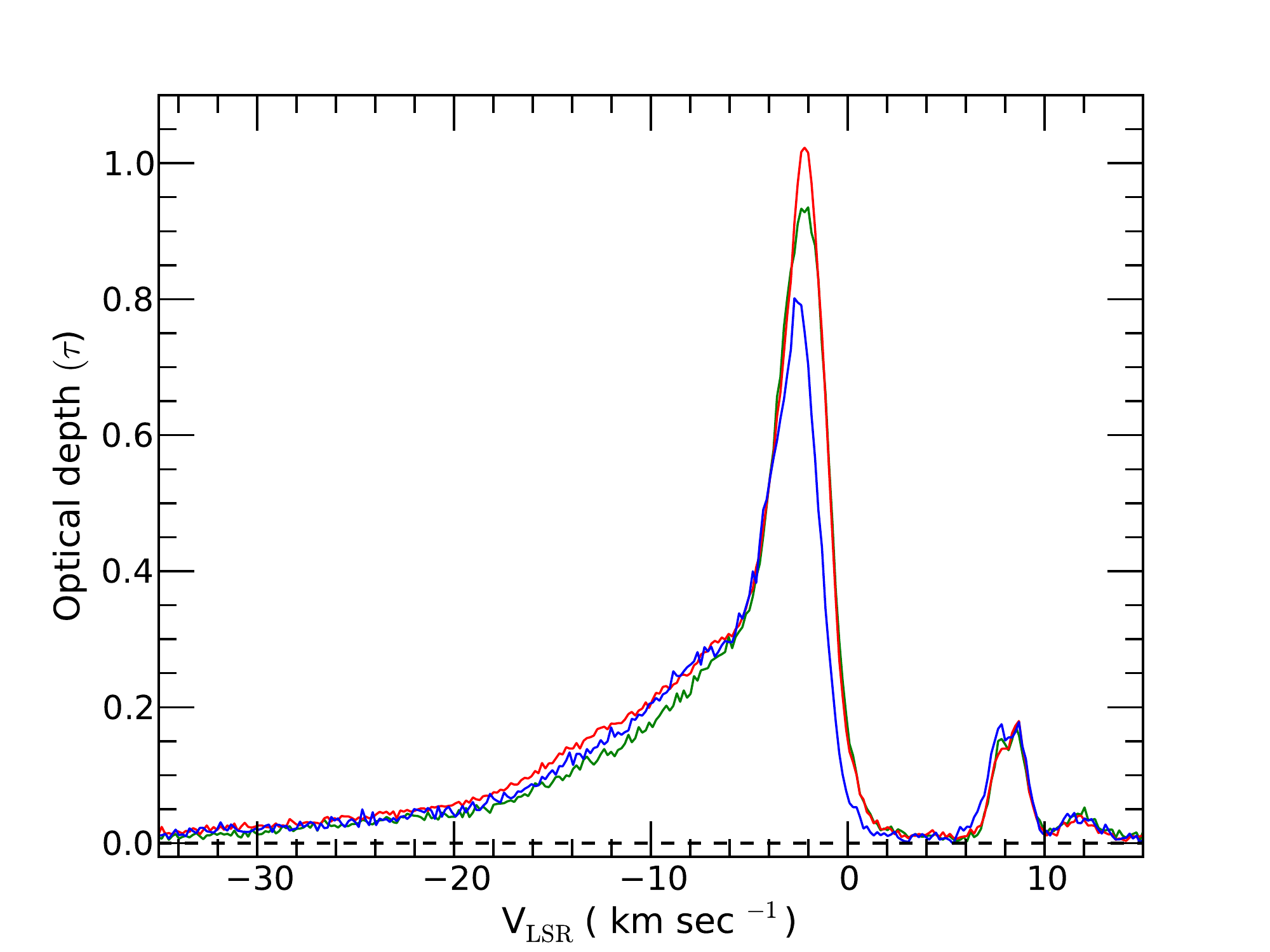}
  \caption{OH 1667~MHz optical depth profile toward the three DR~21~(M) positions (also marked in Fig.~\ref{fig:fig3}), namely, toward the peak of the continuum emission (in red), the region where the magnetic field is significantly detected (in green), and an offset position toward the left of the continuum peak (in blue).} 
\label{fig:fig7}
\end{figure}

The OH optical depth spectra in different part of the DR~21~(M) region has slight variation. In Fig.~\ref{fig:fig3}, three positions, the peak of the continuum, the position corresponding to the most significant OH Zeeman splitting detection (roughly coinciding with the Herschel 70 $\mu$m peak shown in Fig.~\ref{fig:figo}) on the right, and an offset position on the left, are marked in red, green and white, respectively. We show, in Fig.~\ref{fig:fig7}, the OH 1667 MHz optical depth spectra of the corresponding positions in red, green and blue, respectively. Apart from the slight variation of the peak optical depth, the spectra are consistent with one another. Careful multi-Gaussian decomposition of both the 1665 and 1667 MHz spectra show only a weak indication of variation of the central velocity of the negative velocity wing (outflowing component). However, we are severely limited by the coarse angular resolution of the present observations, and high resolution observation will be valuable to probe any possible systematic variation of the velocity of these different components across the region.

\begin{comment}

\end{comment}

\bsp
\label{lastpage}
\end{document}